\documentclass[11pt,a4paper]{article}

\usepackage{amsmath,amssymb}
\usepackage{epsfig,graphicx}
\usepackage{subfigure}
\usepackage{graphicx}
\usepackage{rotating}
\usepackage{cancel}
\usepackage{bm}
\usepackage{color}
\usepackage{comment}
\usepackage{cite}
\usepackage{multicol}
\usepackage{url}

\usepackage
[
colorlinks=true,
citecolor=blue,
urlcolor=blue,
linkcolor=black
]
 {hyperref}

\begin{document}
\numberwithin{equation}{section}
\title{{\normalsize  \mbox{}\hfill DESY Report 20-070}\\
\vspace{2.5cm} 
\Large{\textbf{Straightening of Superconducting HERA Dipoles\\
 for the Any-Light-Particle-Search Experiment\\ ALPS II
\vspace{0.5cm}}}}

\author{Clemens Albrecht$ $, Serena Barbanotti$ $,
 Heiko Hintz$ $, Kai Jensch$ $, Ronald Klos$ $,\\ Wolfgang Maschmann$ $,
 Olaf Sawlanski$ $, Matthias Stolper$ $,
Dieter Trines\thanks{corresponding author: \url{dieter.trines@desy.de}}$ $\\[2ex]
 Deutsches Elektronen-Synchrotron  (DESY)\\
Notkestr. 85, 22607 Hamburg, Germany}


\maketitle

\begin{abstract}
\noindent
 At DESY the ALPS II experiment is being installed in the HERA tunnel to search for axion like particles. A laser beam will be injected into the magnetic field of a string of superconducting dipole magnets, available from the HERA proton storage ring, to produce axion like particles. After passing a light tight wall, the ALPs can reconvert into photons in a second string of HERA s.c.~dipoles.
The sensitivity of the experiment will be increased by two mode-matched optical cavities before and behind the wall.
 The dipoles for the HERA storage ring are curved, suited for stored proton beam.  However, the curvature of the magnets limits the aperture and hence the performance of the optical resonators beyond a certain length. As the sensitivity of the search scales with the length of the magnetic field, the aperture for the optical resonators inside the HERA dipoles was increased by straightening the curved magnet yoke. The procedure of straightening the s.c.~HERA dipoles is described in this report.\\

\end{abstract}
\newpage
\tableofcontents

\section{Introduction }
\label{sec:Introduction}

One of the most important open questions in physics is the nature of dark matter. Among the
candidates from particle physics are axions~\cite{Peccei:2012mb} and other very weakly interacting slim particles (WISPs), e.g. axion-like particles (ALPs) (see~\cite{Igor Irastorzi and Xavier Redondo} for an overview). 
 Light-shining-through-a-wall experiments~\cite{von Bibber:1987mb} allow to search for these particles in the laboratory.
  
More than 10 years ago an experiment of this kind was performed at DESY (ALPS I)~\cite{Ehret:2009sa} making use of a single spare superconducting HERA dipole. This experiment could not  prove the existence of ALPs, but only establish limits on the coupling of ALPs to photons (see below). These limits could be tightened by the OSQAR light-shining-through-a-wall experiment at CERN using two spare LHC dipoles~\cite{OSQAR}. 

 Now  
 the ALPS II \footnote{The ALPS II Collaboration: Albert-Einstein-Institute, Hannover, Germany; Cardiff University, UK; DESY Hamburg, Germany; Johannes Gutenberg-University Mainz, Germany; University of Florida, Gainesville, USA  }
experiment~\cite{Baehre:2009sa} 
is being set up in a straight section of the HERA tunnel with substantially increased sensitivity compared to the previous experiments, using 24  superconducting HERA dipoles. The magnets,
almost 10 m long with a magnetic field of  5.3 Tesla\cite{Wolff:1985mb}, 
 ~are available from the HERA proton storage ring~\cite{HERA:1981mb} not in operation any more since 2007. In addition to a straight section of the HERA tunnel,  infrastructure like cryogenics~\cite{Horlitz:1984mb}, and power supplies needed to operate superconducting magnets are available.

\section{The principal concept of the experiment}
\label{sec:The principal concept of the experiment}

\begin{figure}[htb]
\begin{center}
\includegraphics[width=0.7\textwidth]{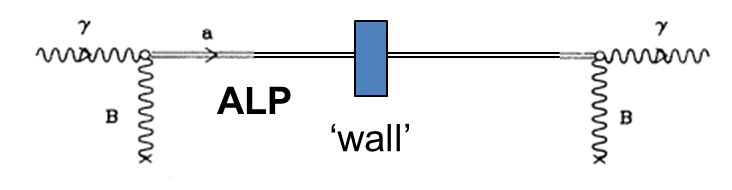}
\caption{\em \small{Physics principle of the experiment. Laser photons interact with the magnetic field B to create an axion like particle. The ALP penetrates the material wall and occasionally reconverts to a photon in the magnetic field behind the wall, to be detected by a photon detector.}}
\label{fig:alps}
\end{center}
\end{figure}

Axions and axion like particles (called ALPs in the following) can be produced by photons traveling in a section of magnetic field, transverse to the flight direction. 
~Similarly ALPs can convert into photons again in another section of the transversal magnetic field. To separate these photons from the photons which produce the ALPs, a  light tight wall is inserted between the two sections of the magnetic field. ALPs penetrate the wall practically without any attenuation (see Fig.~\ref{fig:alps}).

\begin{figure}[htb]
\centering
\includegraphics[scale=0.6]{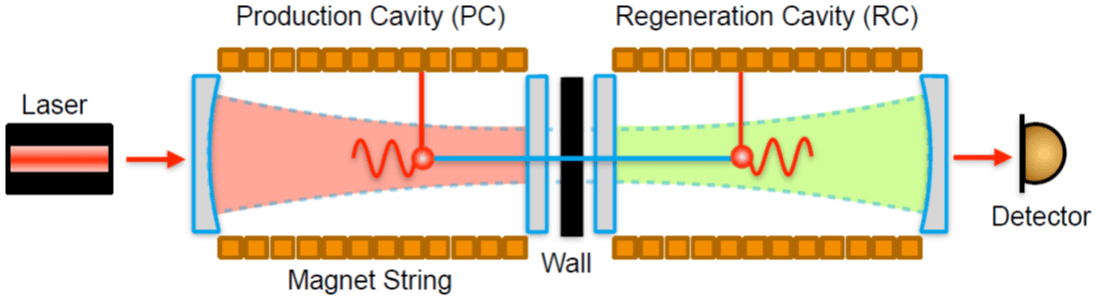}
\caption{\em \small{Schematic layout of the experiment. The probability for the generation of an ALP by a photon gets increased in an optical resonator. Behind the light tight wall a second optical resonator increases the probability of re-conversion to a photon in the magnetic field. (Figure from Aaron~Spector)}
} 
\label{fig:Schematics}
\end{figure}

Laser light is injected into the first section of the magnetic field. To increase the production probability of ALPs, the light is reflected back and forth many times in an optical resonator. The conversion probability of ALPs to light in the second magnetic field section is enhanced also by an optical resonator (enhanced 'spontaneous emission')~\cite{Hoogeveen:1991sa},~\cite{Sikivie:2007sa},~\cite{Fukuda:1996sa} (see Fig.~\ref{fig:Schematics}). For details of the optics layout of the ALPS II experiment see~\cite{Pold:2019sgm} and~\cite{Spector:2016ymd}.

The sensitivity, in measuring the coupling of photons to ALPs, scales with the strength of the magnetic field B in the two sections and their length L, i.e.~with B*L (for details see~\cite{Arias:2010mb}). At DESY strong superconducting dipoles  (see Fig.~\ref{fig:Table}) are available from the 6.3 km long lepton-proton-collider ring HERA. Some of these dipoles will be used to set up two long straight strings of magnets for the ALPS II experiment in a straight section of the HERA tunnel. The geometry of the tunnel limits the number of dipoles to be installed as straight strings to  2*12, i.e. to a length of $\approx$  2* 120 m .

\begin{figure}[htb]
\centering
\includegraphics[scale=0.6]{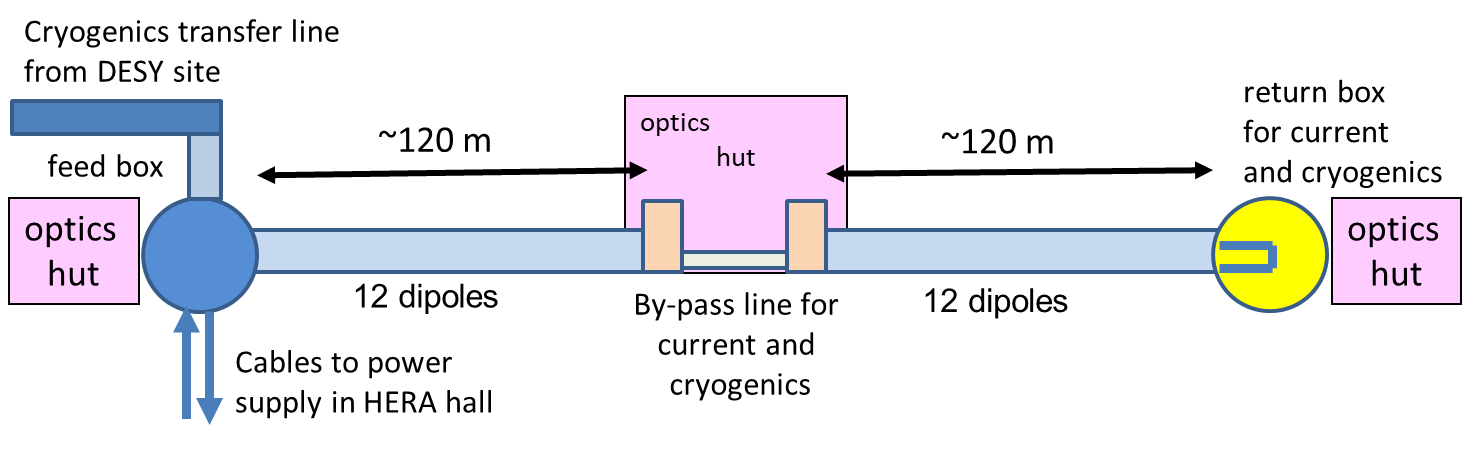}
\caption{\em \small{Schematics of the experimental setup of ALPS II in a straight section of the HERA tunnel. The two strings of 12 straightened dipoles  each are connected by a vacuum insulated bypass line for cold Helium and the magnet current, leaving sufficient space for the setup of the optics in the middle of the experiment. The Helium and the current for the magnet strings are supplied at the box closest to the DESY site.  }
} 
\label{fig:Gesamtaufbau}
\end{figure}

To obtain high quality factors in the optical resonators, leading to high production and regeneration probabilities, a loss of photons on the wall of the vacuum pipe, within the dipoles of each string, must be avoided, i.e. remain below $ 2*10^{-6} $ ~\cite{Poeld:2019sa}. This requires that the aperture of the vacuum pipe must be larger than the envelope of the photon beam within each string of dipoles.  A loss of photons by scattering off gas molecules is negligible as the optical resonators are located in the high vacuum of the beam pipe.

Unfortunately, the vacuum pipes within the HERA dipoles\cite{Trines:1985mb},\cite{Boehnert:1992mb}  are curved, suited for the storage of protons. For light however, moving on a straight line, the horizontal aperture  in the optical resonators is reduced from 55.3 mm (inner diameter of the beam pipe) to about 35 mm only. This aperture allows only for 2*40m of magnetic length, i.e.~2 strings of 4 HERA dipoles each, without limiting the performance of the optical resonators. 

The aperture can be increased by straightening the HERA dipoles leading to a practically loss free length of about 120 m for each resonator, which matches the above mentioned maximum length, given by the tunnel geometry, for installing straight strings of dipoles. The ALPS II setup  with 2*12 straightened dipoles in the HERA tunnel is shown schematically in Fig.~\ref{fig:Gesamtaufbau}.
 
\section{The superconducting  HERA dipole} 

\begin{figure}[tbp]
\centering
\includegraphics[width=0.8\textwidth]{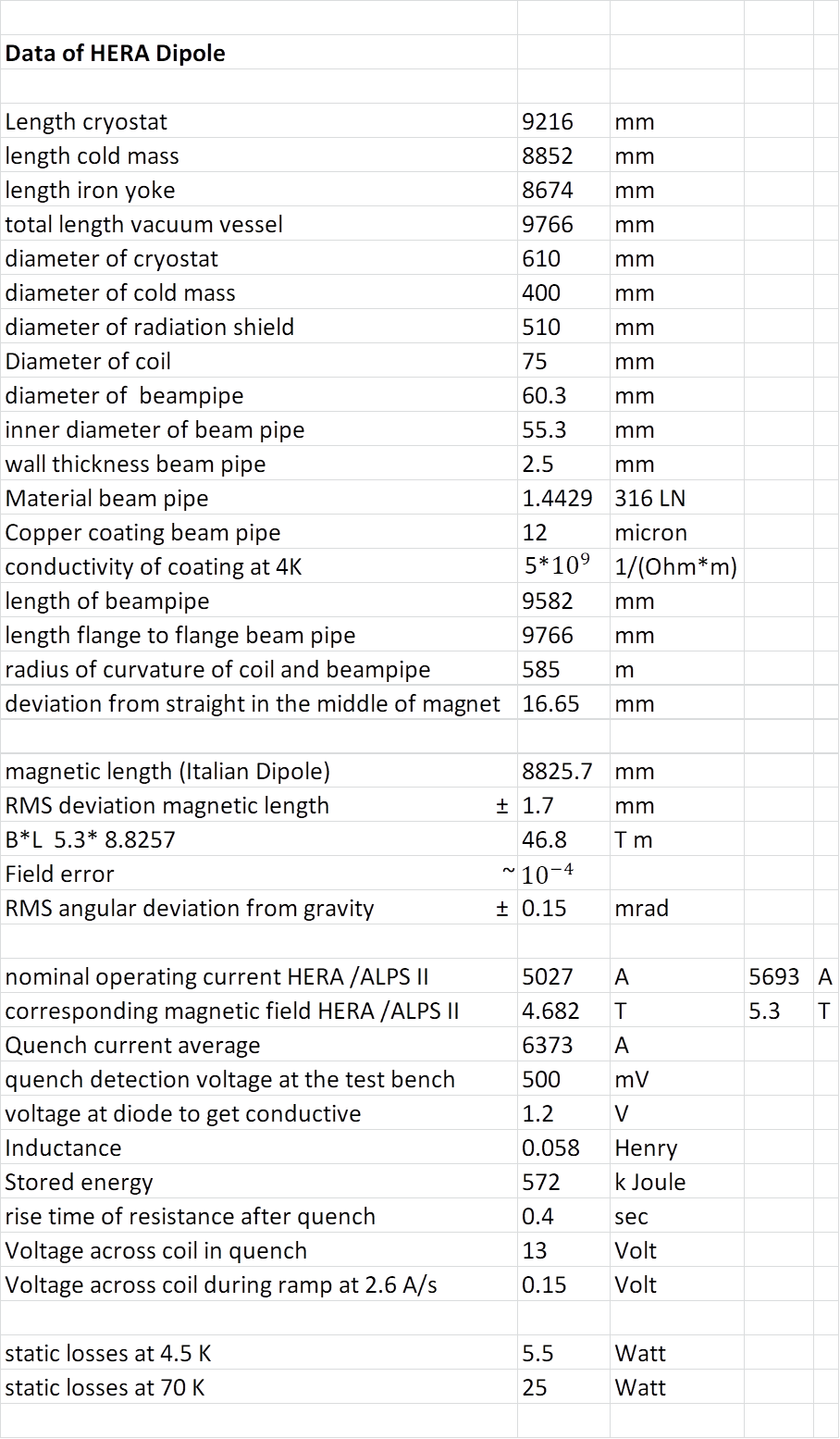}
\caption{\em Table of HERA dipole properties.} 
\label{fig:Table}
\end{figure}

\begin{figure}[h]
\centering
\includegraphics[scale=0.5]{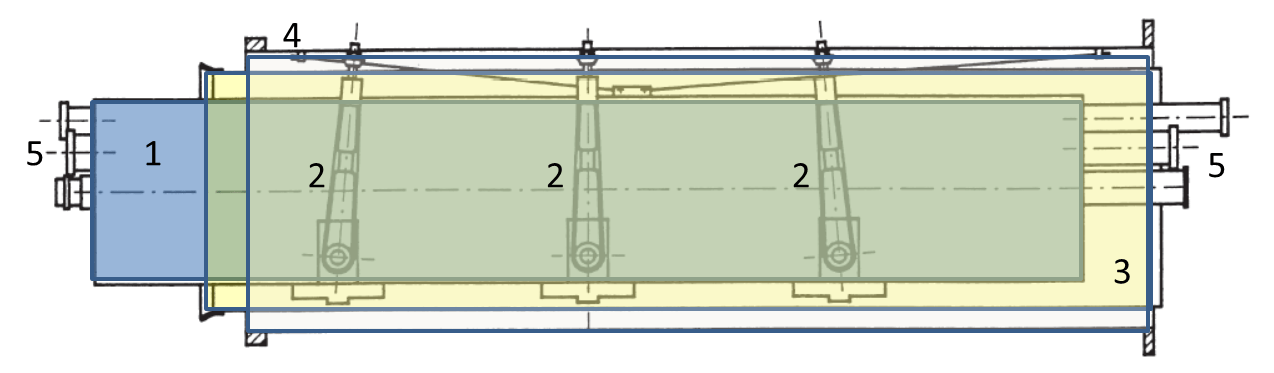}
\includegraphics[scale=0.5]{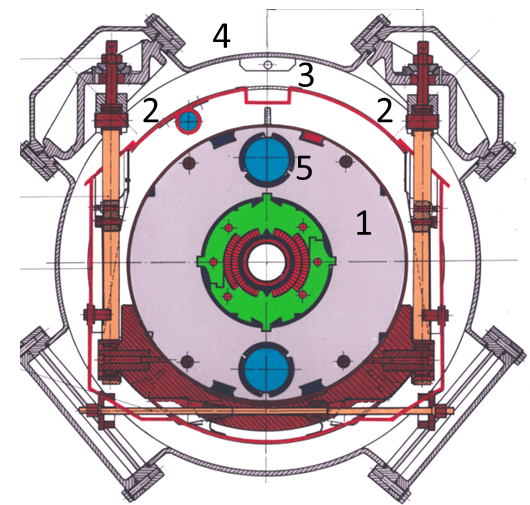}
\caption{\em \small{Schematic view of the HERA dipole. 1 Helium vessel containing cold mass, 2 Suspension, 3 Radiation shield, 4 Vacuum vessel, 5 Helium pipes.}} 
\label{fig:lengthcut}
\end{figure}

The 'cold mass' of the HERA dipole, i.e.~the superconducting coil and the magnet iron,  is supported from the vacuum vessel at three planes along the cryostat (see Fig.~\ref{fig:lengthcut}). The suspensions (6 in total) also support the radiation shield. 

As indicated in  Fig.~\ref{fig:lengthcut}, the cold mass is contained in a stainless steel vessel (Helium vessel) welded from two half cylinders. The beam pipe in the middle of the cold mass is surrounded by the magnet coil wound from superconducting cable ('Rutherford' cable).
The coil is held by strong clamps counteracting the magnetic forces. The magnetic field outside the coil is concentrated in slices of magnet iron surrounding the clamps .

Some of the properties of the HERA dipole are compiled in  Fig.~\ref{fig:Table}, including results from measurements of the magnetic field \cite{Meinke:1991}.

\begin{figure}[h]
\centering
\includegraphics[scale=0.4]{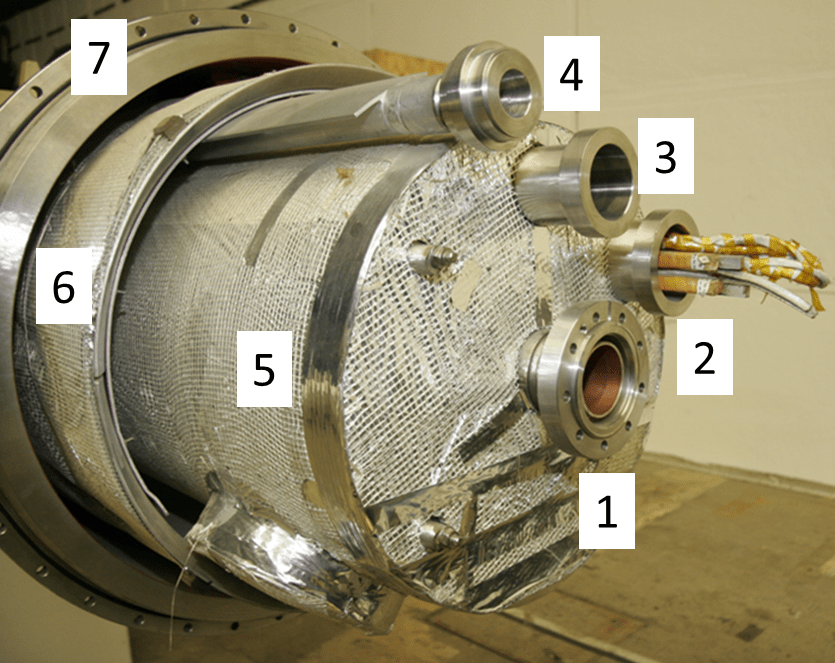}
\includegraphics[scale=0.34]{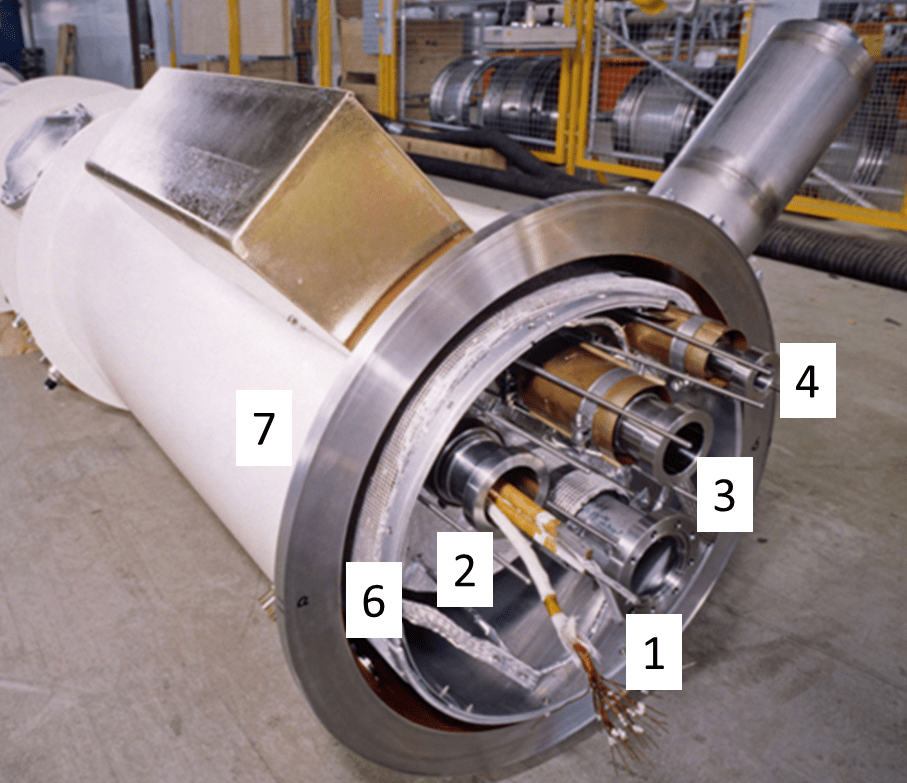}
\caption{\em \small{View of the two ends of a HERA dipole.
 1: Beam tube, 2: One phase Helium pipe with superconducting cables for the magnet coil, 3: Two phase Helium pipe, 4: Helium gas pipe for the cooling of the radiation shield,  5: Helium vessel containing the cold mass, 6: Radiation shield, 7: Vacuum vessel.}} 
\label{fig:endview}

\end{figure}

 Figure~\ref{fig:endview} shows the ends of a HERA dipole with the pipes for 1- and 2-phase Helium emerging from the  Helium vessel. The s.c. cables to and from the coils in the magnet are routed through the 1-phase tubes at the ends of the Helium vessel. The radiation shield is cooled by a single pipe by Helium gas at 40-70K.

\begin{figure}[tbp]
\centering
\includegraphics[scale=0.4]{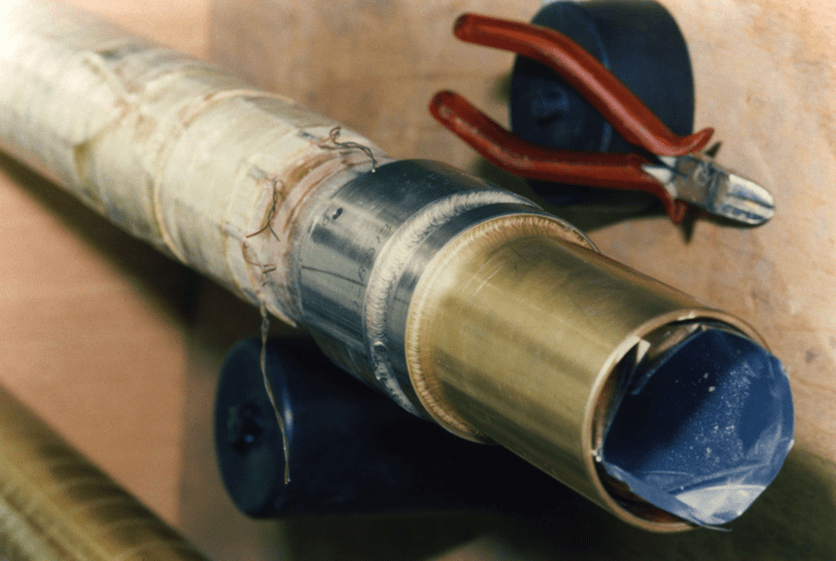}
\includegraphics[scale=0.36]{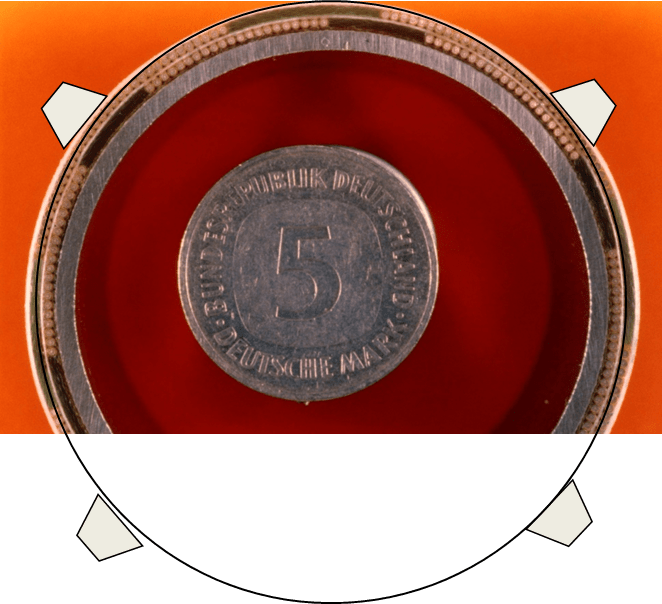}
\caption{\em \small{The beam tube of the HERA dipole. Indicated are (right) the Vespel spacers which keep the pipe centered in the coil and force the pipe to follow the deformation of the yoke.}} 
\label{fig:beamtube}
\end{figure}

Originally the cold mass was fabricated as a straight unit. By a strong press the two half cylinders of the Helium vessel were forced around the magnet iron and bent to the required radius before joining the half cylinders by welding\footnote{The dipoles were fabricated by companies in Italy (Ansaldo and Zanon) and Germany (ABB).}. The beam pipe (see Fig.~\ref{fig:beamtube}) was forced to follow the curvature by spacers, glued to the outside of the pipe. The outer vacuum vessel  forms a polygon, roughly following the curvature of the cold mass (see Fig.~\ref{fig: komplett} right).

The cold mass could, in principle, be straightened again by cutting the welds on the Helium vessel. However this would require a complete disassembly of the cryostat and the proper tooling, which mostly would have to be rebuilt. 

As this would be a very costly procedure, we chose a cheaper method to increase the aperture of the beam pipe without disassembly of the cryostat: bending the cold mass inside the vacuum vessel by 'brute force' (see section:~The straightening procedure)\footnote{A similar method was considered independently at CERN by P.Pugnat for LHC dipoles, as mentioned in\cite{Sikivie:2007sa}.}.

There was serious concern, that the superconducting coils might be damaged by this way of straightening. To ensure that the magnets could still be operated at the design current for ALPS II of 5963 A, all magnets were tested after straightening up to their respective quench currents on a test bench available at DESY.

\section{The test bench}
\begin{figure}[h]
\centering
\includegraphics[scale=0.6]{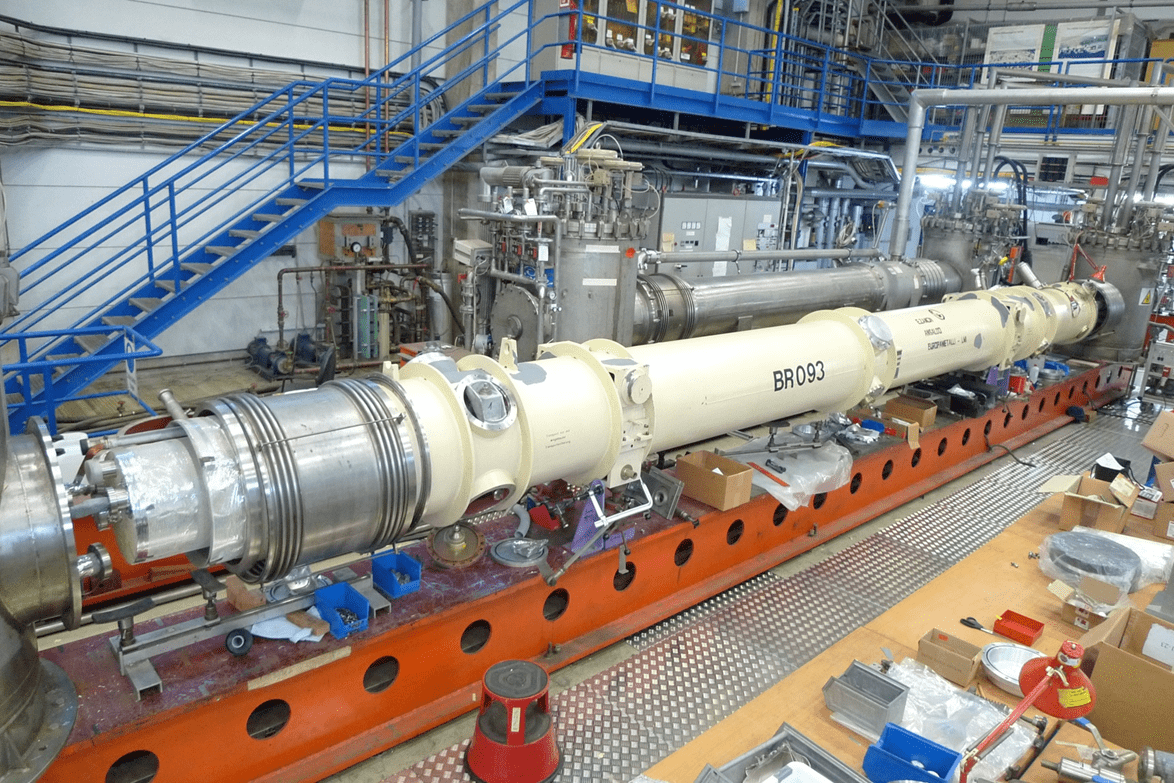}
\caption{\em \small{Remaining bench for cryogenic tests of HERA dipoles with a dipole on the bench. Note the open sliding sleeve connection of the vacuum vessel to the cryogenics end box. }} 
\label{fig:Testbench2}
\end{figure}

One dipole test bench\footnote{The test bench was used already for the ALPS I experiment.} is still available from the magnet test facility on the DESY site~\cite{Barton:1987mb} for the lepton-proton-collider HERA (see Fig.~\ref{fig:Testbench2}), with all the cryogenics tubing, current leads, valves, sensors, safety systems, and so on. Helium, liquid and gas, can be supplied by the existing vacuum insulated transfer line from the central DESY cryogenics plant. 

The original power supply of the test facility, capable to supply up to 7000 A at 20 Volts, is available. The old quench detection and protection system~\cite{Mess:1987mb} for magnet operation was updated.
 A turbo pump station and a roughing pump for the insulating vacuum are available. So, everything required for a test of HERA dipoles is operational. 

After a dipole was positioned on the test bench, the electrical connection between the coil and the power supply was established by soldering the superconducting cables emerging from the dipole cryostat (see Fig. \ref{fig:endview}) with the ones from the feed box of the test bench~\cite{Stolper:2018mb}.

 When the survey of the beam pipe during the straightening of the magnet (see section: The survey of the beam pipe) was completed, the Helium lines between the boxes and the magnet were sealed by special Aluminium seals. Copper half shells were connected around the bellows in the Helium pipes, to prevent them from buckling when pressurized. The Helium pipes, and the magnet at the flanges were shielded from thermal radiation with super insulation (see section: The new suspensions) after leak checking the connections of the Helium pipes. Then the  open flanges and the sliding sleeves on the  vacuum tank (see Fig.~\ref{fig:Testbench2})   were closed, to evacuate the cryostat vessel.  At about $ 10^{-4} $ mbar in the cryostat vessel, the cool down of the cold mass to 4K proceeded.

 It should be noted, that the beam tube was open within the cryostat on the test bench, being evacuated together with the  cryostat vessel. This will be different for the magnet strings in ALPS II, where the vacuum pipes of adjacent dipoles will be connected, and the beam tube vacuum will be separated from the insulating vacuum.

\section{The straightening procedure}
\subsection{General}
\begin{figure}[htp]
\centering
\includegraphics[scale=0.5]{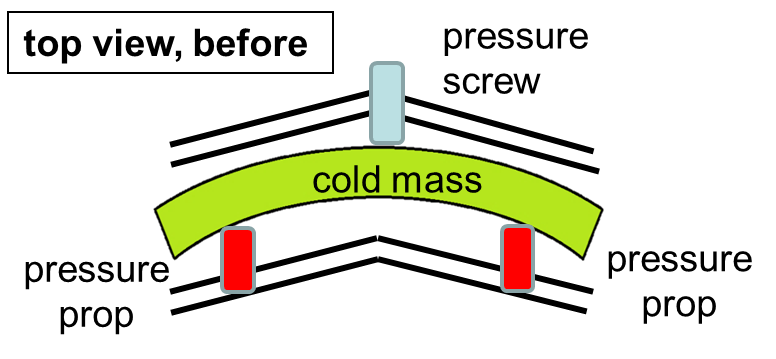}
\includegraphics[scale=0.17]{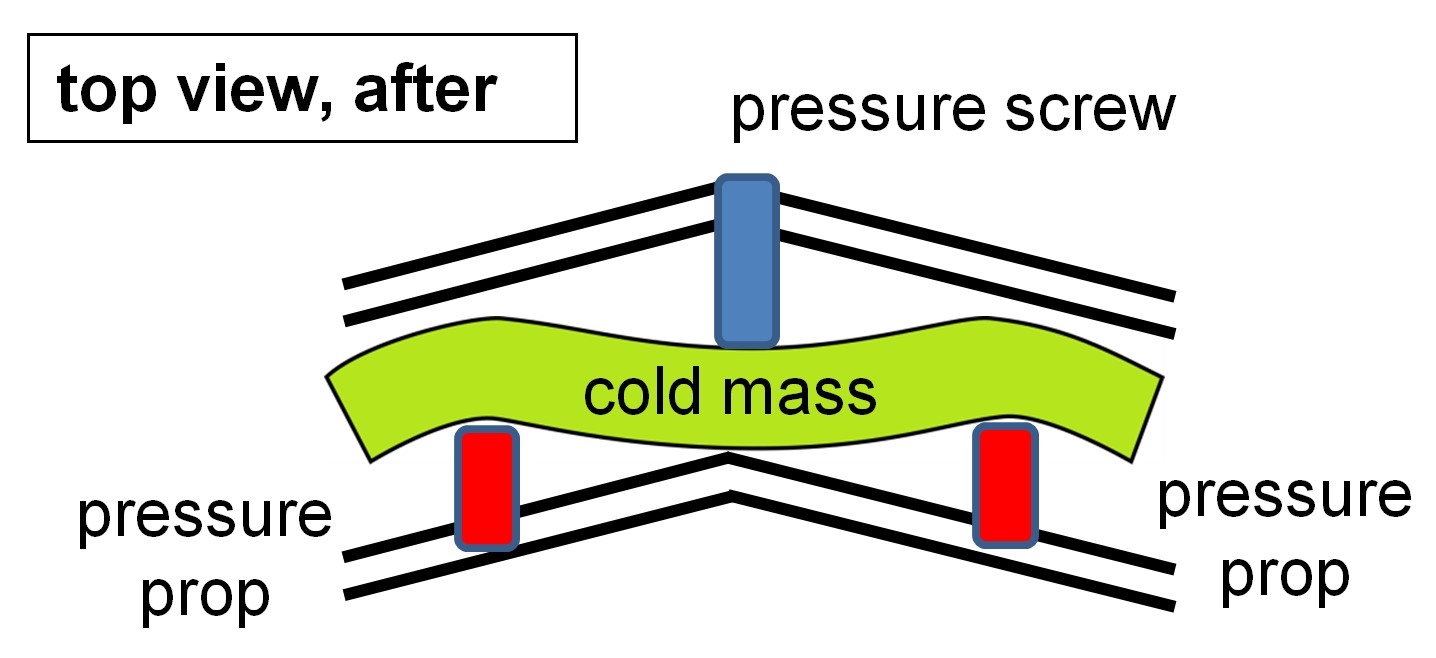}
\caption{\em \small{Schematics of straightening. Left:~Before applying the deforming force, Right: The deformation forces the pipe to develop two 'camel humps,'  exaggerated in the figure for better illustration. This deformation yields the largest achievable horizontal aperture.}} 
\label{fig:before}
\end{figure}

\begin{figure}[htp]
\centering
\includegraphics[scale=0.75]{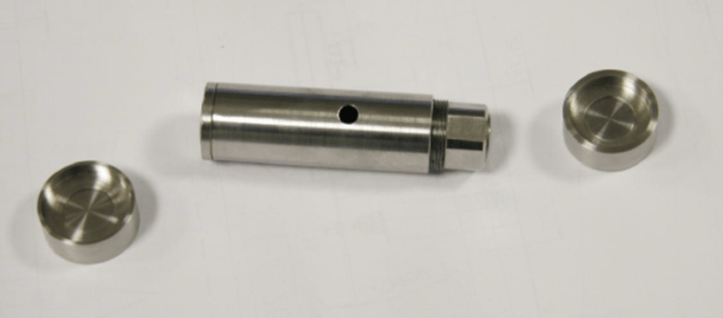}
\includegraphics[scale=0.7]{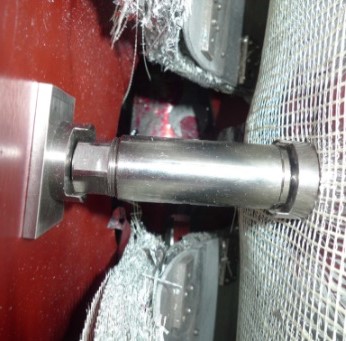}
\caption{\em \small{Outer pressure prop parts (left) and prop inserted into the cryostat (right).} 
\label{fig:pressure prop}}
\end{figure}

\begin{figure}[htp]
\centering
\includegraphics[scale=0.5]{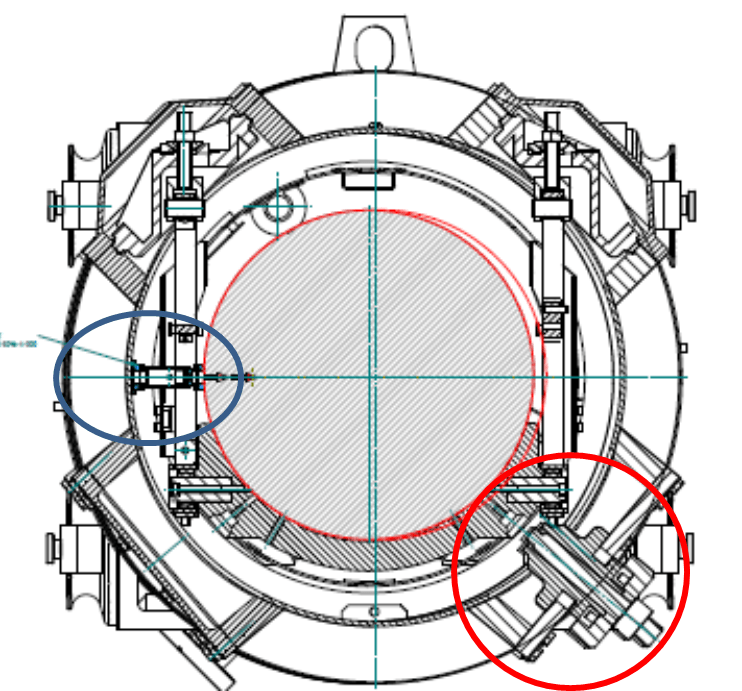}
\includegraphics[scale=0.6]{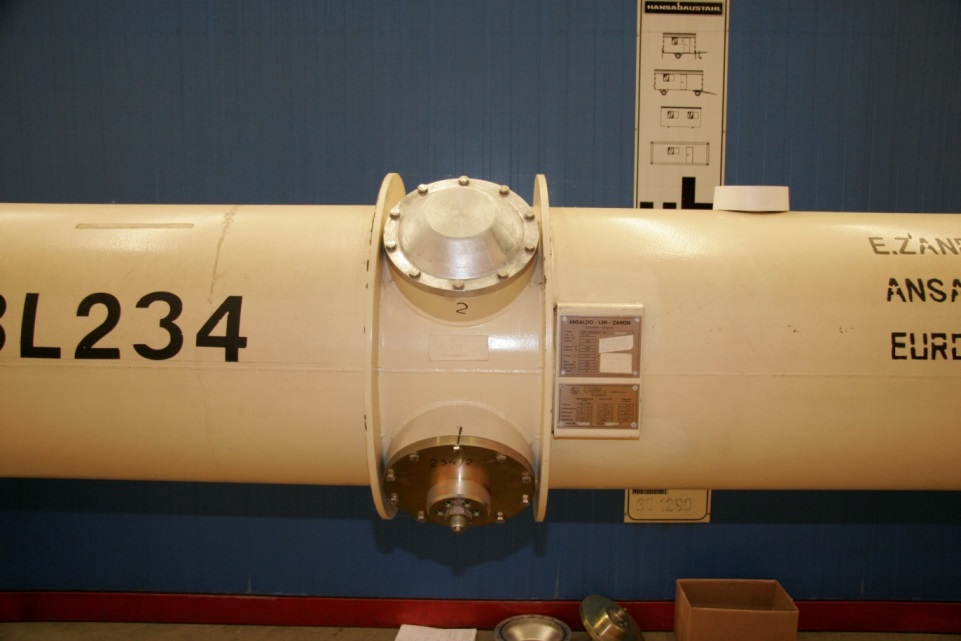}
\caption{\em \small{Left: Cross section of cryostat with inserted outer pressure prop (blue) and transportation fixture (red). Transportation fixture mounted into vacuum vessel of the cryostat (right). }} 
\label{fig:cross2}
\end{figure}

The straightening of the magnet and the vacuum pipe, to increase the horizontal aperture of the beam tube for the optical resonators, was performed on the test bench. It was achieved by fixing the position of the ‘cold mass’ at the outer suspensions of the magnet and pushing (and thus deforming) the cold mass in the middle (see Fig.~\ref{fig:before}). The cold mass  was fixed  by inserting and tightening Titanium pressure props (see Figs.~\ref{fig:pressure prop} and~\ref{fig:propschem})  between the cold mass and the vacuum vessel~\cite{Meyer:2010mb}.  By a transportation fixture (see Fig.~\ref{fig:cross2}) on the opposite side of the cryostat, the cold mass was prevented from moving, while tightening a pressure prop.\footnote{Normally, the transportation fixtures prevent the cold mass from moving with respect to the vacuum vessel during transportation of the magnet.} 

\begin{figure}[htp]
\centering
\includegraphics[scale=0.6]{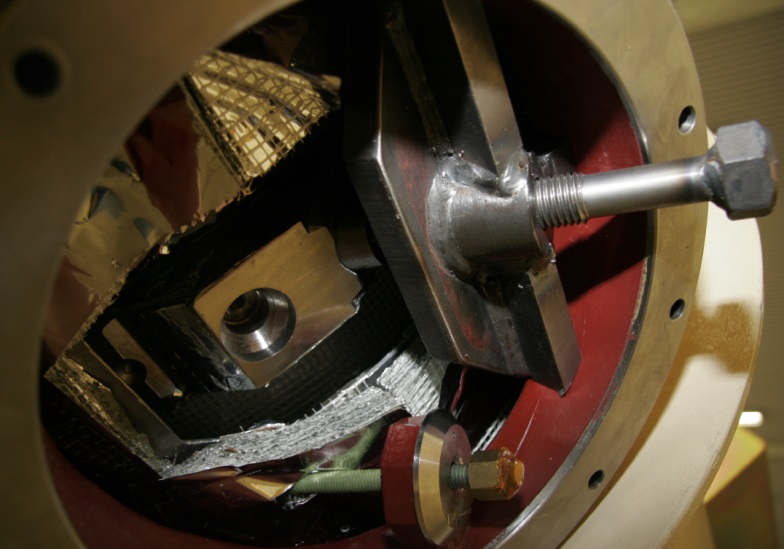}
\includegraphics[scale=0.55]{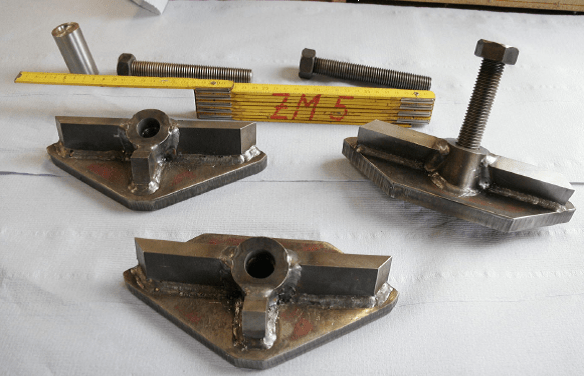}
\caption{\em \small{Pressure screw mounted into the vacuum vessel in the middle of the cryostat for the deformation of the cold mass (left), Pressure screws (right).}} 
\label{fig: Druckschraube}
\end{figure}

The deformation in the middle was achieved by a  steel screw ('pressure screw')  (see Fig.~\ref{fig: Druckschraube}) exerting a force of about 40 kN, inserted between the vacuum tank and the Helium vessel at the lower flange of the vacuum vessel. 

\begin{figure}[htp]
\centering
\includegraphics[scale=0.5]{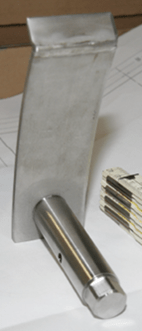}
\includegraphics[scale=0.5]{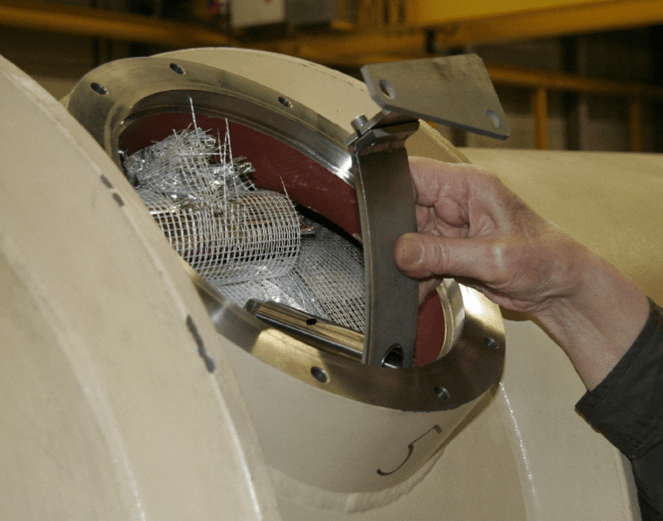}
\includegraphics[scale=0.7]{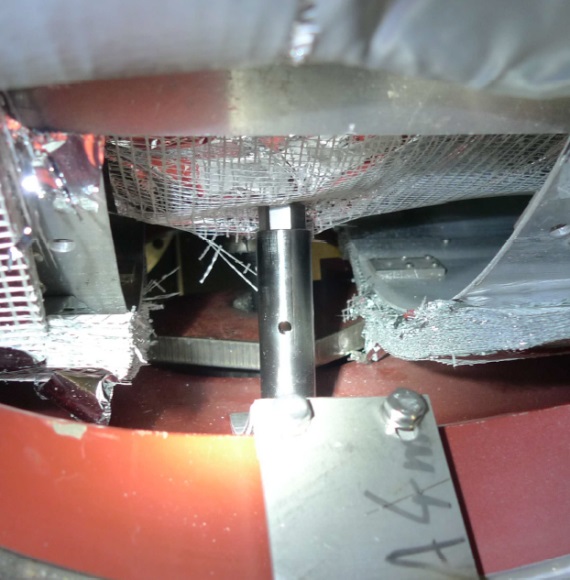}
\caption{\em \small {Left:~Middle pressure prop, Middle:~Inserting the prop, Right:~Inserted pressure prop in the middle of the cryostat}} 
\label{fig: propmiddle}
\end{figure}

\begin{figure}[htb]
\centering
\includegraphics[scale=0.6]{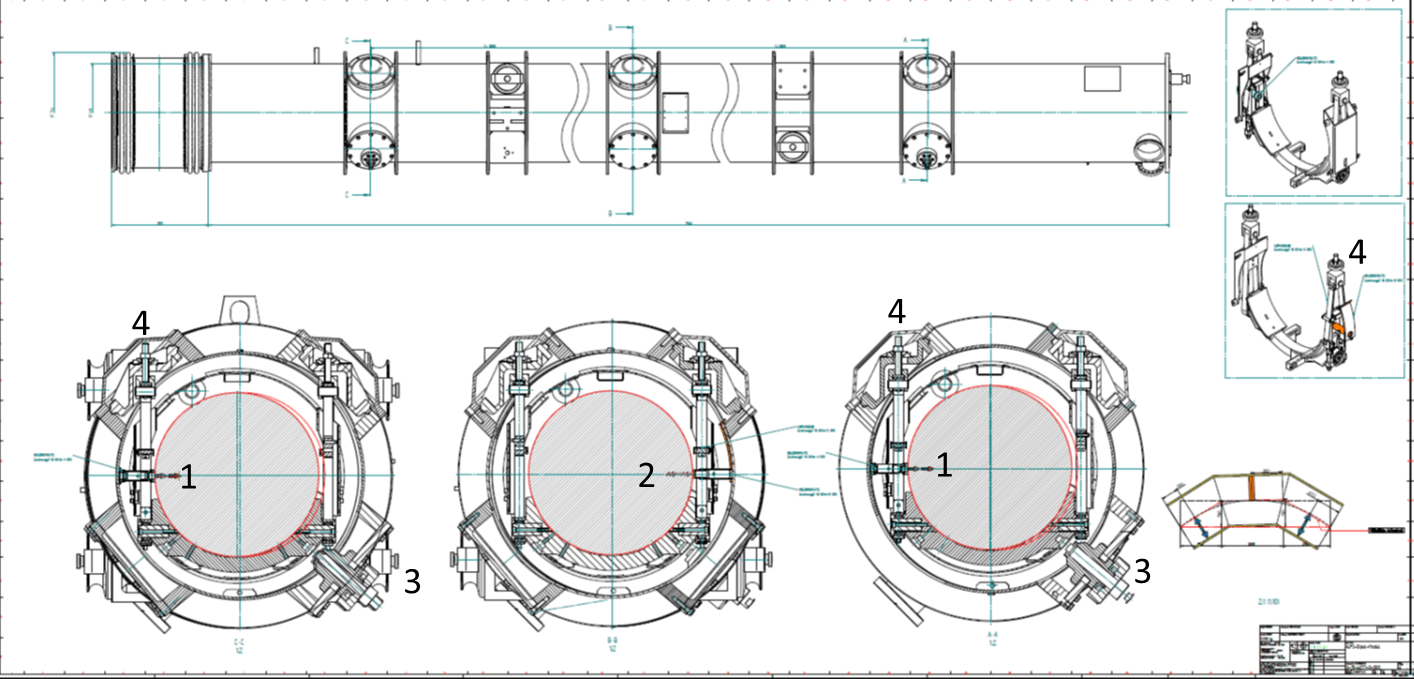}
\caption{\em \small{Schematic view of the cryostat after the straightening. 1: Outer pressure prop, 2: Middle pressure prop, 3: Transportation fixture, 4: Suspension.} } 
\label{fig: komplett}
\end{figure}

When the aimed for deformation was achieved, another Titanium pressure prop was inserted (see Fig.~\ref{fig: propmiddle}) from the upper flange above the pressure screw. After tightening the pressure prop, the pressure screw and the transportation fixtures were removed. With all three pressure props installed and tightened, the deformation of the yoke is maintained by the tension of the yoke, counteracted by the  vacuum tank via the pressure props. Fig.~\ref{fig: komplett} shows the cryostat with three Titanium pressure props inserted before the removal of the transportation fixtures.

The ends of the beam tube cannot be straightened independently, as no force can be applied beyond the outer suspension planes. The ends just rotate around the fix points, given by the outer pressure props, due to the bending of the beam tube in the middle (see Fig.~\ref{fig:before}). The middle of the cold mass was bent in a way, to force the beam  tube to develop two 'camel humps' (see Fig.~\ref{fig:before} and also Fig.~\ref{fig: Verformung}). This deformation yields the largest achievable horizontal aperture~\cite{Meyer:2010mb}.  In the figure the deformation is exaggerated for better illustration.

 The pressure prop in the middle (see section: The pressure props) of the cryostat constitutes a fix point for the thermal shrinkage/expansion during cool down/warm up of the cold mass.

 Finite element calculations, determining the additional stress on the Helium vessel by the straightening procedure, were performed~\cite{Meissner:2015mb} and presented to the agency for pressure vessel safety (TUEV). The calculations showed that all stresses are well below the limits set by pressure vessel regulations.

There is a detailed report  on the work required for the straightening~\cite{Hintz:2018mb}, partially presented in a poster on the PATRAS workshop 2018~\cite{Albrecht:2018mb}.


\subsection{The survey of the vacuum pipe}
\begin{figure}[htp]
\centering
\includegraphics[scale=0.8]{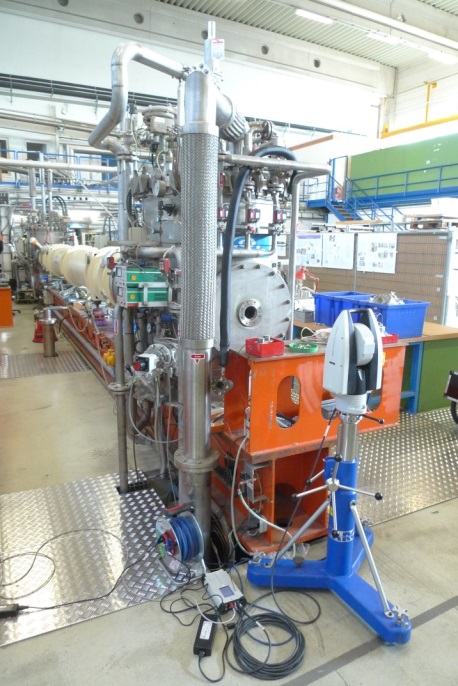}
\includegraphics[scale=0.8]{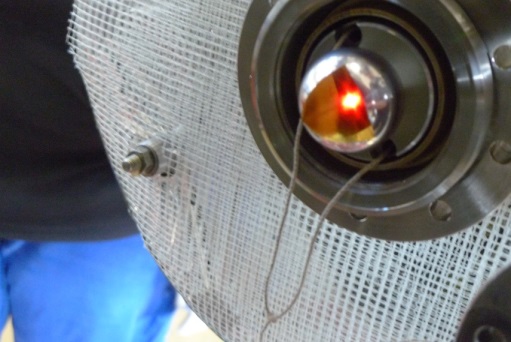}
\includegraphics[scale=0.6]{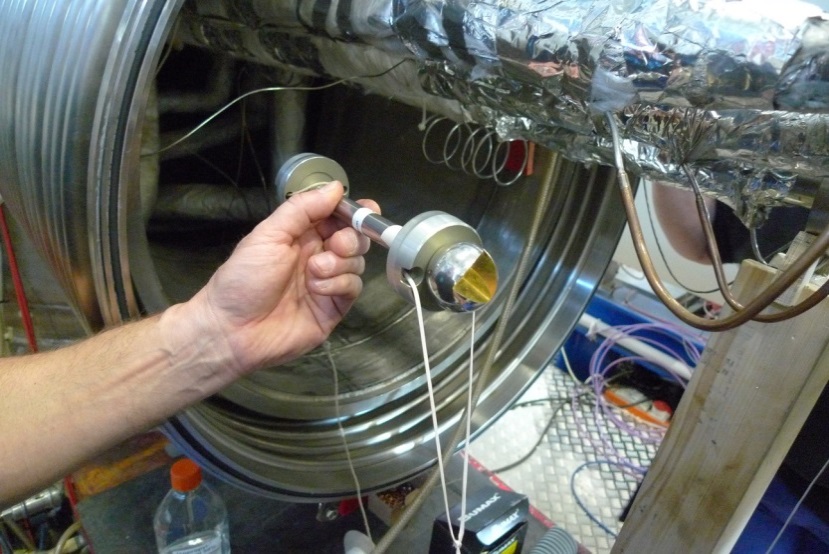}
\caption{\em \small {The laser tracker to measure the position of a reflector in the vacuum pipe of the dipole through an open flange in the end box. Middle: reflector mouse  inserted into the beam pipe; Right: before insertion. The string connected to the mouse is used to pull the reflector through the beam pipe.} } 
\label{fig:tracker}
\end{figure}

The position of the center of the beam pipe before, during, and after the deformation was measured by the DESY survey group with a laser tracker and a so-called mouse with a reflector attached. The mouse was pulled by a string through the vacuum pipe along the length of the magnet, while the laser tracker continuously measured the position of the reflector (see Fig.~\ref{fig:tracker}) through an open flange in the middle of the end box of the test bench.

\begin{figure}[htb]
\centering
\includegraphics[scale=0.4]{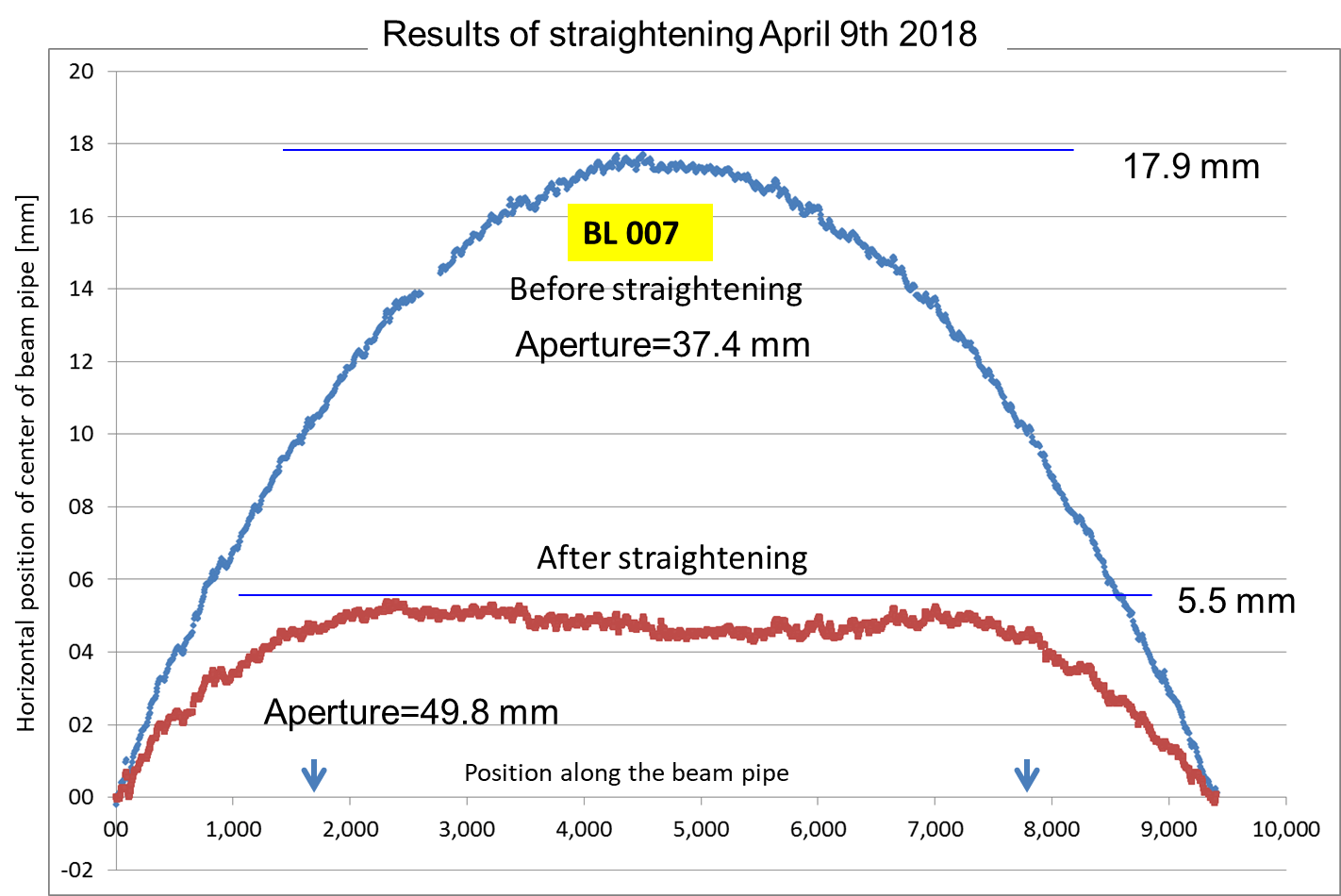}
\caption{\em \small{The result of  straightening a dipole. Shown is the measured center of the vacuum pipe before and after straightening. The positions of the outer pressure props are indicated by arrows. The maximum achievable aperture by the deformation is about 50 mm. The inner diameter of the beam tube is 55.3mm.} } 
\label{fig: Verformung}
\end{figure}

 A typical result of straightening a dipole vacuum tube is shown in Fig.~\ref{fig: Verformung}, in comparison with the original curved shape of the vacuum pipe. The success of the deformation was judged by the horizontal aperture achieved.

The result of the survey of the beam pipe, i.e. the position of the beam pipe center line after straightening, was transferred to marks, welded to  the outside of the vacuum vessel. When setting up the straight magnet strings for ALPS II in the HERA tunnel, these survey marks will allow to align  the dipoles to yield the largest possible overall horizontal aperture within the strings.

The curvature of the beam pipe before the straightening slightly varied among the dipoles. The maximum deviation from the straight line connecting the ends of the beam tube (17.9 mm in the example shown in Fig.~\ref{fig: Verformung}) varied by $\pm$~2.5 mm between the extremes of 16 and 21 mm. The average was 18.4 mm. There is a correlation between this maximum deviation and the achieved aperture after the straightening procedure. In general, the larger the achieved final aperture, the smaller is the deviation from a straight line connecting the ends of the original beam tube.

The beam tubes showed also deviations from the horizontal plane, i.e.~in vertical direction by $\pm$ 1 to 2 mm, both before and after straightening. These deviations had no influence on the achieved horizontal apertures.

During one cryogenic test, the mouse with reflector was installed into the vacuum pipe in the middle of the dipole,  to monitor the position of the vacuum pipe after cool down of the dipole. The position of the reflector in the cold vacuum pipe could be measured with the laser tracker through a quartz window on a flange in the end box of the test bench.

The measurement showed that the horizontal aperture increase, achieved by the deformation of the cold mass, is reduced by the cool down of the magnet by about 0.5 millimeter, caused by the thermal shrinkage of the pressure props and the cold mass. The impact on the performance of the two 120 m long optical resonators for ALPS II is tolerable (see section: Results).

To check whether the straightening of the dipole would suffer from transportation between DESY and the HERA hall East for installation into the tunnel, a straightened dipole (BR 221) was put on a transport trailer and driven from the DESY site on public roads  to the HERA hall East and back. Then the center line of the beam tube was measured again. 
It was identical to the one measured before the transportation,

\subsection{The pressure props}

\begin{figure}[htb]
\centering
\includegraphics[scale=0.3]{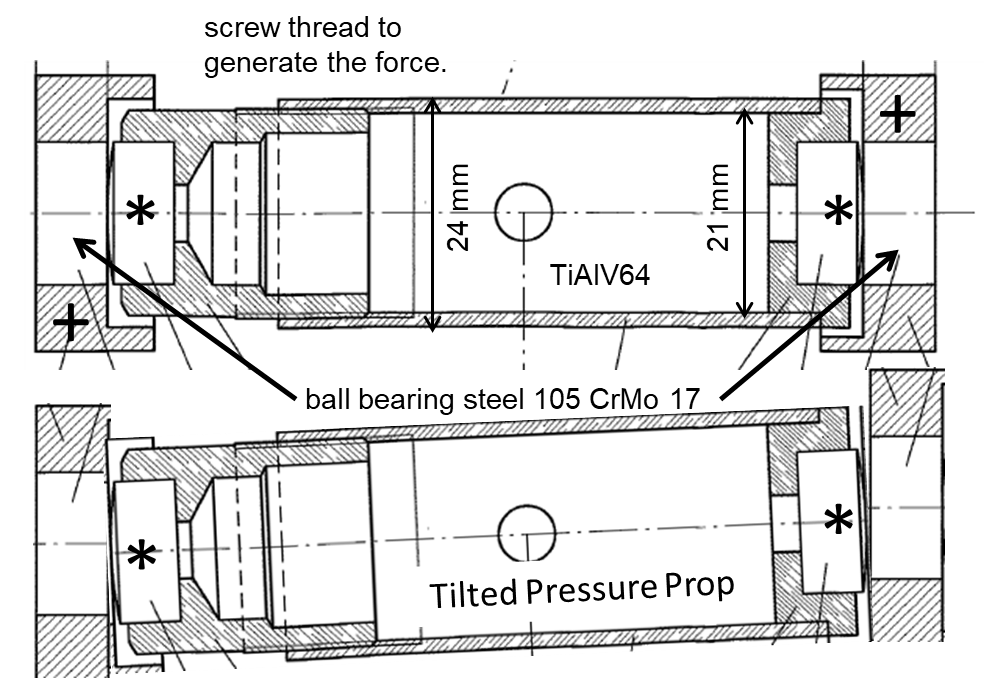}
\includegraphics[scale=0.3]{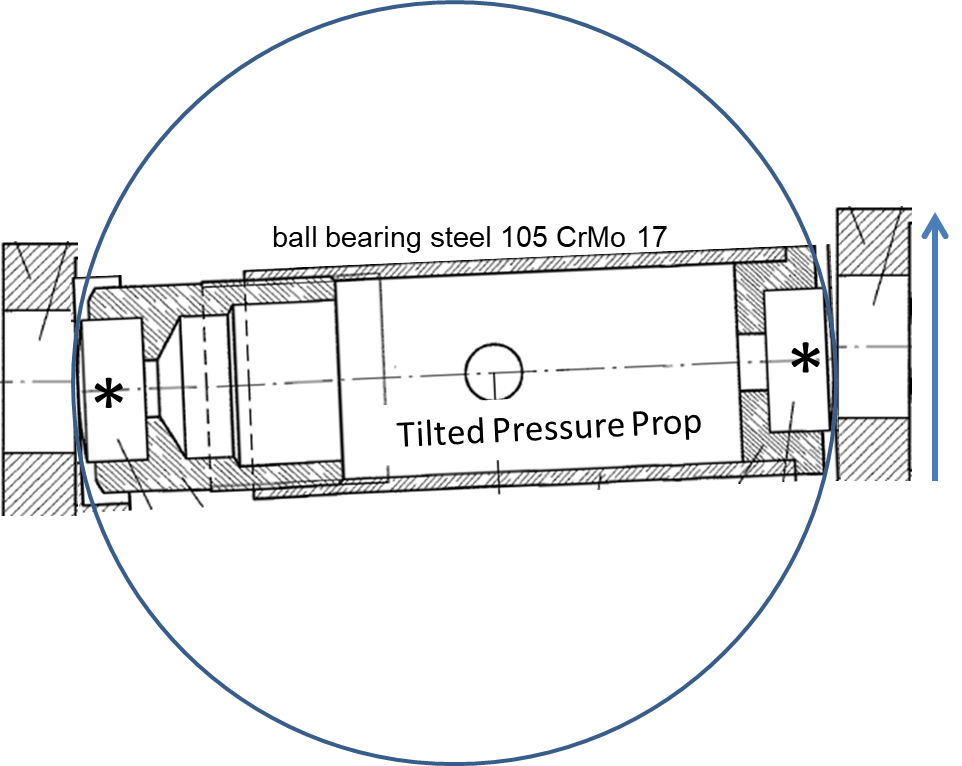}
\caption{\em \small{Schematic view of the pressure prop at an outer suspension in the cryostat with the thin walled Titanium tube. Note the spherical sections at both ends (marked by *) from ball bearing steel. The full sphere for the roll of the pressure prop is indicated on the right. The support cup at the vacuum vessel stays fixed, while the cup on the Helium vessel moves with the shrinkage/expansion of the Helium vessel. The support cups are marked by +.}} 
\label{fig:propschem}
\end{figure}

As the deformation of the yoke is elastic, the deforming force by the pressure props must be maintained, also at cryogenic temperatures.  However, the pressure props constitute a thermal short between the Helium vessel at 4K and the vacuum vessel at room temperature. To keep the additional heat flow to the 4K level of the cryogenics supply as low as possible, the props were made from a thin walled (1.5 mm) Titanium tube (see Figs.~\ref{fig:pressure prop},~\ref{fig:cross2},~\ref{fig: propmiddle}, and~\ref{fig:propschem}),  a material with low thermal conductivity, comparably low thermal expansion, and large mechanical strength. In addition, the contact area at both temperature levels is small for the props at the outer suspensions, as the ends of the props form sections of a sphere. The thermal flux from room temperature to the yoke at liquid Helium temperature was estimated to about 1 Watt per prop (see section: Results).

The props near the ends of the dipole must allow for the length change of the yoke with respect to the vacuum vessel ($\approx$ 30 mm total) during cool down and warm up and yet maintain the deforming force. Shaping the ends of the props as sections of a sphere, allows for a tilt of the props during cool down or warm up (see Fig.~\ref{fig:propschem}), without changing the distance between the vacuum vessel and the cold mass~\cite{Meyer:2010mb}, except for the thermal shrinkage of the props and the cold mass (see above). 

The deforming forces are always perpendicular to the surfaces of the support cups, (see Fig.~\ref{fig:propschem} and Fig.~\ref{fig:pressure prop}) resting on the Helium and the vacuum vessel, thus reducing the momentum on the prop and the danger of it tipping over. The support cups of the outer pressure props at the vacuum vessel stay fixed, while the cups  on the Helium vessel move with the shrinkage/expansion of the Helium vessel.

The Titanium tube of the pressure prop in the middle was screwed into a stainless steel strip (5 mm thick and 50 mm wide), which matches the inner surface of the vacuum vessel (see Fig.~\ref{fig: propmiddle}). The steel strip remained in the cryostat after straightening  the yoke.
The middle pressure prop has a flat contact area to the Helium vessel and does not tilt during thermal cycles. It thus constitutes a fix point for the thermal shrinkage/expansion during cool down/warm up of the cold mass.

\begin{figure}[htp]
\centering
\includegraphics[scale=0.25]{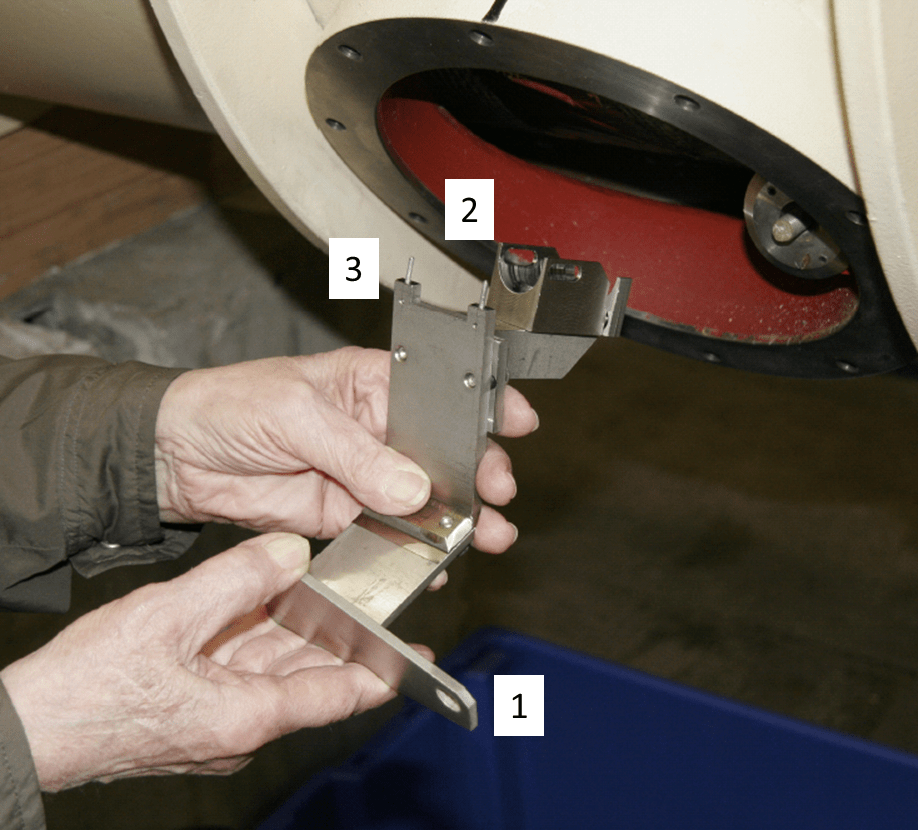}
\includegraphics[scale=0.6]{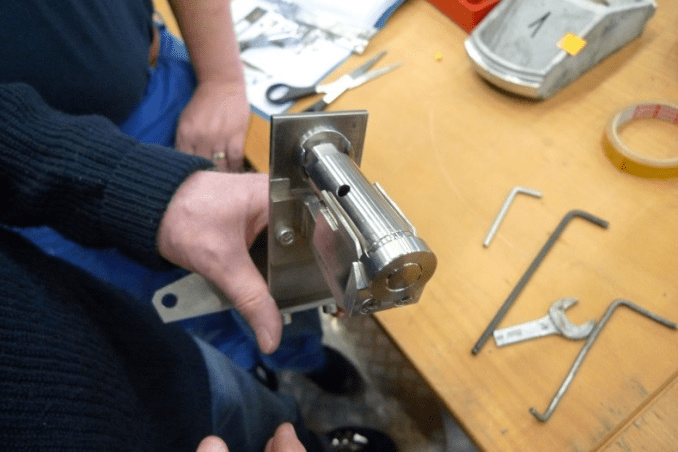}
\includegraphics[scale=0.8]{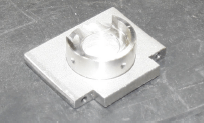}
\caption{\em \small{Tool for proper placing of the outer pressure props. Left:~1:~Plate to fasten the tool to the flange of the vacuum vessel, 2:~Cradle to hold the pressure prop, 3:~Pins  to connect to the outer pressure plate.  Middle: Complete prop in the assembly tool. Right:~Outer pressure plate with support cup. }} 
\label{fig:insertiontool}
\end{figure}

To place the pressure props at the right position and angle within the cryostat, special mounting tools were developed (see Fig.~\ref{fig:insertiontool} and also Fig.~\ref{fig: propmiddle}). After positioning the prop, the assembly tool was removed. It should be noted, that all props are finally held in position inside the cryostat only by the tension between the  vacuum vessel and the Helium vessel.

The proper choice of materials and the concept were validated by a test of an outer pressure prop in vacuum at liquid Nitrogen temperature. A force of 40 kN was exerted on the prop in a test device while moving one side back and forth by $\pm$15 mm several hundred times. No problem with the prop was encountered.
 
\subsection{The new suspensions}

\begin{figure}[h]
\centering
\includegraphics[scale=0.47]{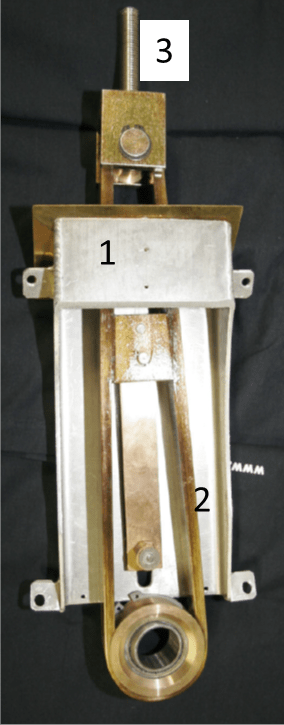}
\includegraphics[scale=0.5]{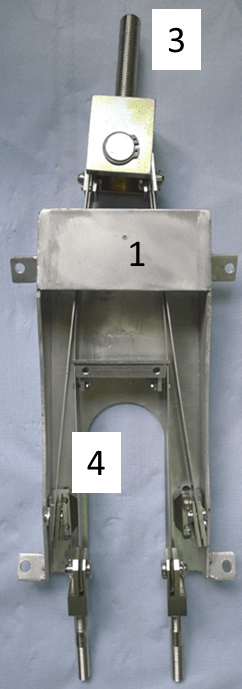}
\includegraphics[scale=0.5]{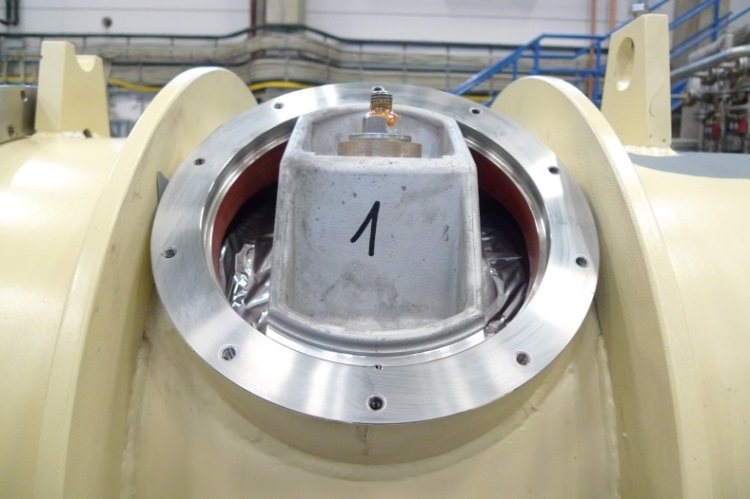}
\caption{\em \small {Original suspension of the cold mass removed from the cryostat (left).  The suspension with the radiation shield box (1) prevented the installation of a pressure prop into the cryostat (see text). The new suspension, with a slit machined into the radiation shield box and Titanium strips replacing the G10 loop, allowed to slide the suspension over the installed pressure prop (middle).~1:~radiation shield box, 2:~G10 supporting loop, 3:~suspension screw, 4:~Titanium support strips. Right: Suspension screw at top of the suspension }  } 
\label{fig:originalsuspension}
\end{figure}

\begin{figure}[htp]

\centering
\includegraphics[scale=0.44]{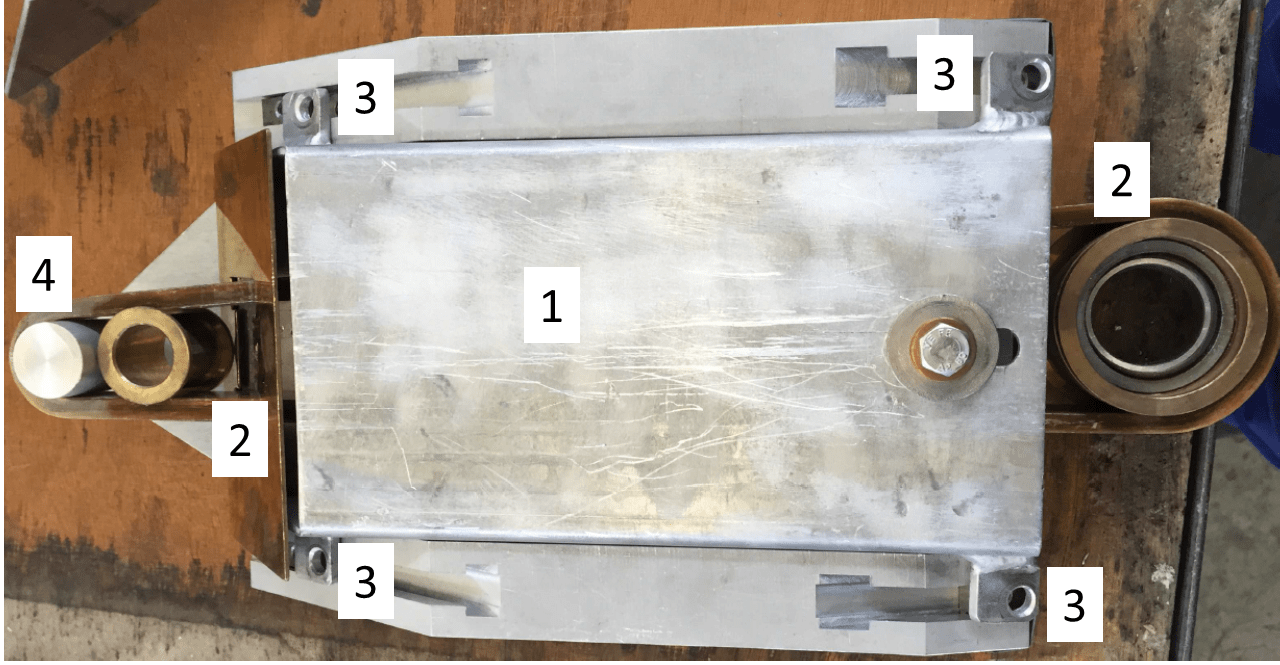} \includegraphics[scale=0.4]{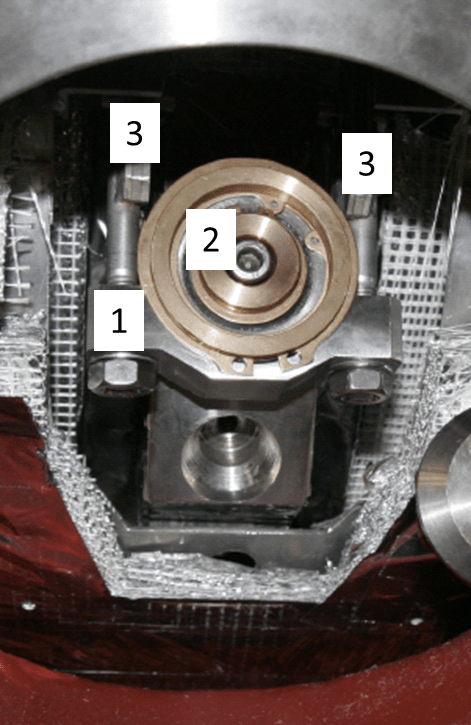}
\caption{(Left) \em \small{Original suspension of cold mass mounted to transfer tool. 1: Radiation shield box, 2: G10 supporting loop, 3: Fastening ears, 4: Transfer pivot.}   } 
\label{fig:transfertool}
\caption{(Right)\em \small{ Steel bracket fastened to the lower end of the new suspension replacing the support by the G10 loop. 1: Bracket, 2: Support pivot, 3: Lower end of new suspension} } 
\label{fig:bracket}

\end{figure}

The insertion of  pressure props at any of the three positions, foreseen in the midplane of the magnet, was impossible with  the original suspensions in place. The radiation shield box, connected to the suspension, was blocking the insertion of a prop  (see Fig.~\ref{fig:originalsuspension} and  Fig.~\ref{fig:cross2}). Therefore the original suspensions of the cold mass with the radiation shield box had to be removed from the cryostat for the insertion of a pressure prop.

 In the beginnings of the straightening studies, which were performed on a defective dipole used as an exhibit, all three suspensions were removed at the same time. The cold mass was still safely suspended by the remaining three suspensions. However, the yoke bend vertically by several millimeters, which took away part of the aperture and was difficult to restore due to the large forces required. 

Therefore, in the final procedure, applied to all straightened dipoles, only one suspension was removed at a time, according to the following sequence: First, one of the outer suspensions was removed, then the corresponding pressure prop was installed. After a modification of the suspension (see below)  it was reinstalled, now carrying again the weight of the cold mass. Then followed the same steps for the other outer suspension. Finally, the suspension in the middle was removed and modified. After the straightening of the magnet,  the  middle suspension was reinstalled in the same way as the other suspensions.
The vertical deformation was small for this procedure (a few tenth of a millimeter) and could easily be restored by the new suspensions (see  below). 

Obviously, an installed pressure prop prevented the re-installation of the original suspension. Therefore a slit was machined into the radiation shield box, to allow for the re-installation of the suspension, when the pressure prop was in place. Also the G10 loops, supporting the cold mass, were replaced by an open structure, realized by Titanium strips (10 mm wide and 1 mm thick) to pass by the inserted pressure prop (see  Fig.~\ref{fig:originalsuspension}). The thermal flow to 4K due to the modified suspension was increased by 0.3 Watt per suspension i.e. 0.9 Watt per dipole (see section: Results).

The free length of the support screws (see  Fig.~\ref{fig:originalsuspension} right) at the top of the suspensions were measured before their removal  from the cryostat.
 Before removing the shield box from an original suspension,  the positions of the ears -- for the fastening screws to the radiation shield -- were marked on a special tool, while the upper part of the G10 loop had tight contact with the transfer pivot of the tool (see  Fig.~\ref{fig:transfertool}). After machining, the shield box was aligned in the device to the marks, before fastening it to the new suspension, positioned by the transfer pivot of the tool.
These measures ensured, that the ears of the radiation shield box were at the same position on the new suspension as on the original one with respect to the support screw. 

 After the new suspension was inserted into the cryostat, the shield box was screwed to the radiation shield. An especially tailored package of super insulation foils, sliding over the pressure prop, was attached to the shield box. Then the length of the support screw on top of the suspension was adjusted to its original -- before measured -- value, ensuring that the new suspension -- and thus the radiation shield -- was at the same position as the original one. Finally a support bracket was fastened by nuts to the suspension. The bracket, now carrying the cold mass (see  Fig.~\ref{fig:bracket}), was pushed up by nuts with only little force against the support pivot of the cold mass. The nuts were secured by a counter nut.

Finally the lower end of the suspension and the support pivot were shielded against thermal radiation by an Aluminium cap, covered with super insulation foil and attached to the radiation shield by two screws.

\begin{figure}[htp]
\centering
\includegraphics[scale=0.45]{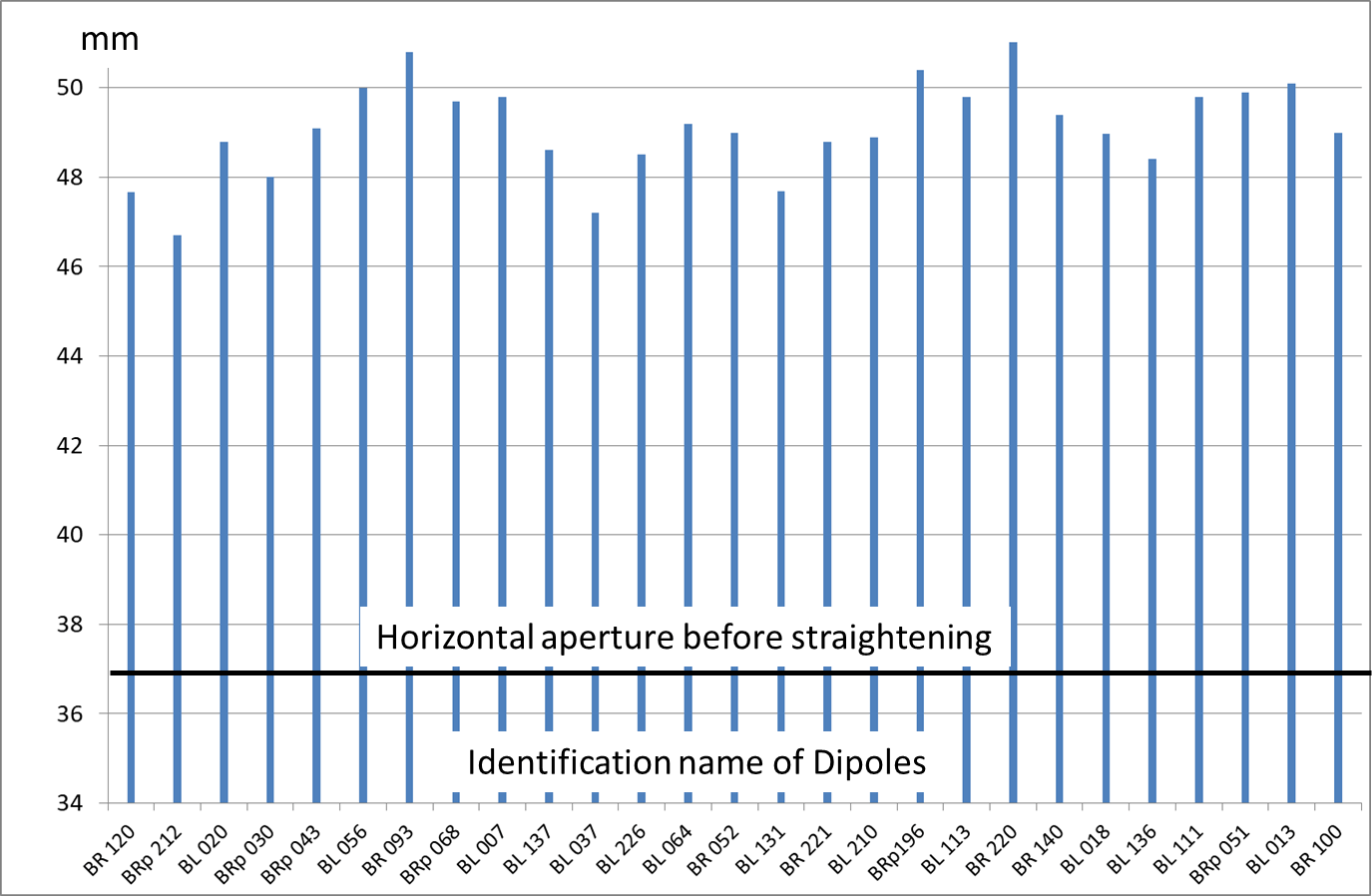}
\caption{\em \small{Achieved horizontal apertures after the straightening of HERA dipoles in mm on the vertical axis versus the identification name of the dipoles. For comparison the average horizontal aperture before the straightening is shown. For photon losses below $ 2*10^{-6} $ an aperture of 45.5 mm is required at the ends of the experiment, for infrared light with 1064 nm wavelength used in ALPS II.} } 
\label{fig:aperture}
\end{figure}

\section{Results}

There were 23 spare magnets  from the HERA proton ring available for straightening and subsequent cryogenic operation on the test bench.
It should be noted, that these magnets were stored in a heated hall on the DESY site for about 25  years after their original test .
All tested dipoles belong to the large contribution from Italy (50\% of all superconducting dipoles) to the HERA project, built by the companies Ansaldo and Zanon.

Two magnets from the spares showed a missing or bad electrical connection, either at 4K or at room temperature, between voltage tabs on the superconducting coils -- needed for quench detection -- and the corresponding feed-through at the cryostat vessel. These magnets were sorted out. During the test of one magnet, the power supply tripped for unknown reasons (no quench). As a precaution, this magnet was also sorted out from installation into the ALPS II setup. Thus, 20 of the spare dipoles remained for the setup of the ALPS II experiment, designed for a total number of 24 dipoles.

The ALPS II experiment requires space for optics clean rooms at the ends of the magnet strings (see Fig. \ref{fig:Gesamtaufbau}).  To gain this space and, at the same time, yielding additional magnets for the ALPS II setup, several dipoles of the HERA ring were removed from the tunnel.  These dipoles, which also belong to the above mentioned Italian HERA contribution, were also straightened and operated at the test bench. 

Fig.~\ref{fig:aperture} shows the obtained horizontal apertures in mm on the vertical axis versus the identification name of the dipoles. For photon losses below $ 2*10^{-6} $ in the optical resonators, for infrared light at a wavelength of 1064 nm, an aperture of 45.5 mm is required at the outer ends of the experiment. The aperture values shown in the figure have to be reduced by 0.5 mm, due to the shrinkage of the pressure props and the Helium vessel during cool down (see section: The survey of the beam pipe).
~For comparison the average horizontal aperture before the straightening of about 37 mm is shown.

\begin{figure}[htp]
\centering
\includegraphics[scale=0.45]{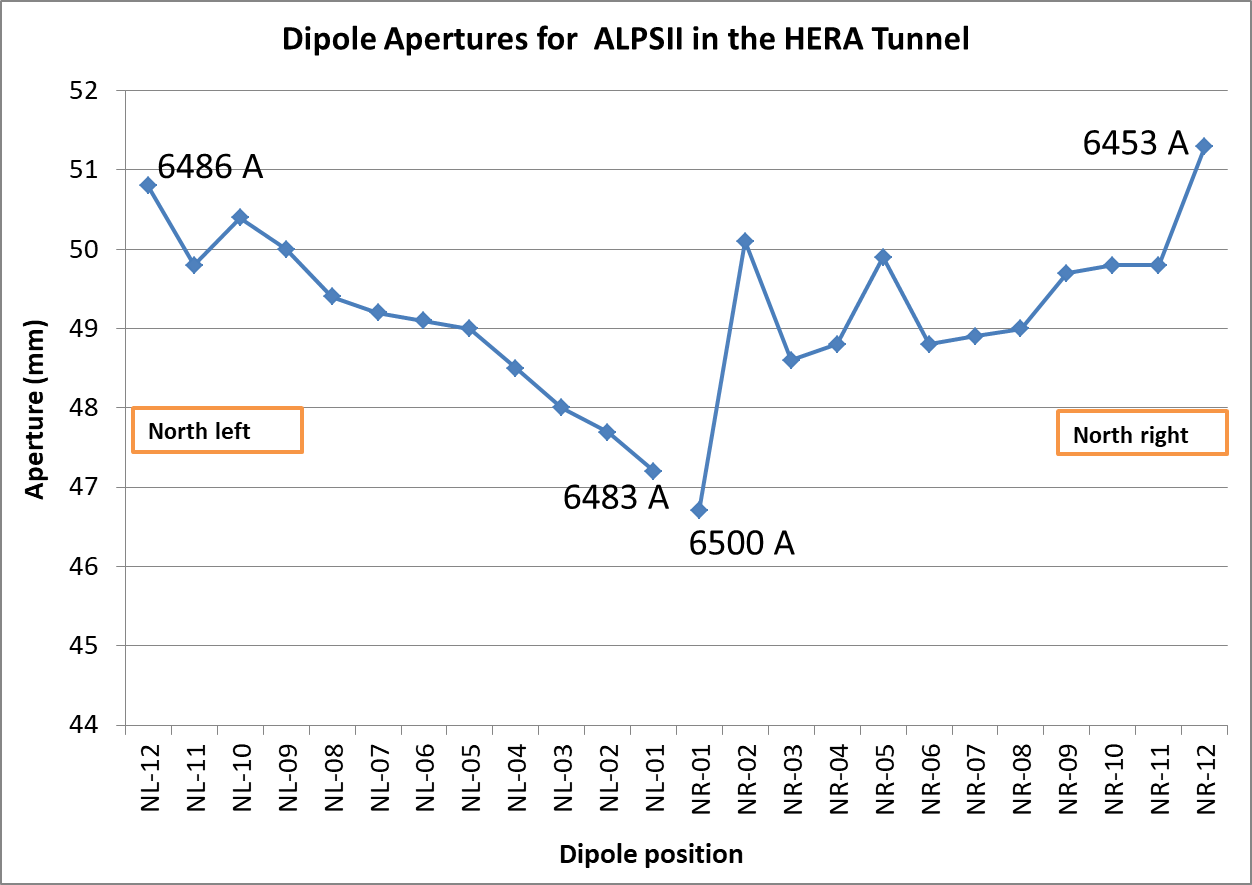}
\caption{\em \small {Horizontal apertures of the dipoles to be installed in the HERA tunnel. Large apertures are at the outer ends of the magnet strings, smaller apertures close to the middle of the experimental setup. The numbers attached at the outer and inner positions show the obtained quench currents.}} 
\label{fig:Aufstellung}
\end{figure}

The cross section of the stored photon beam in the optical cavities is larger at both ends of the experiment than in the middle (see Fig.~\ref{fig:Schematics}).~Therefore, in order to minimize clipping of the stored photon beam in the optical resonator, dipoles with larger apertures were selected to be positioned near the outer ends of the experiment and magnets with smaller aperture near the middle of the experiment.                    Fig.~\ref{fig:Aufstellung}  shows the apertures of the dipoles vs. their selected positions in the ALPS II setup. This selection  leaves about $\pm$ 2mm free space for adjustment uncertainties, without limiting the performance of the optical resonators.

\begin{figure}[htp]
\centering
\includegraphics[scale=0.45]{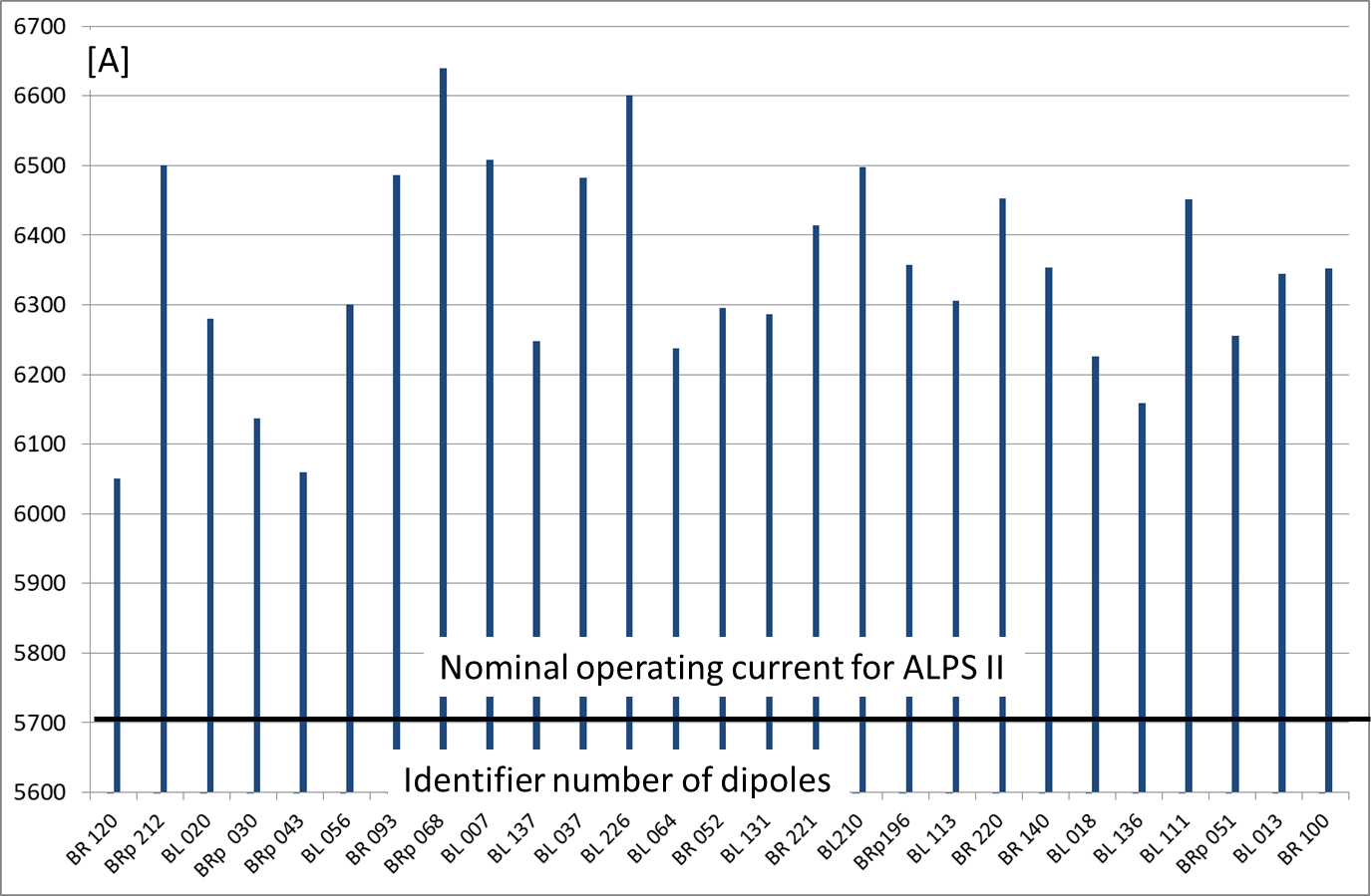}
\caption{\em \small {Measured quench currents of straightened dipoles at the test bench} } 
\label{fig:quench}
\end{figure}

All dipoles were operated on the test bench up to their quench current twice. In addition all dipoles were operated continuously at the nominal  operating current  of the ALPS II experiment for about 8 hours. 
Fig.~\ref{fig:quench} shows that all dipoles have quench currents well above the nominal operating current of 5963 A.

 Near the ends of the strings for ALPS II, magnets with large quench currents ($>$6400 A) were selected, to compensate for potential additional heat loads from the neighboring boxes or the cryogenic bypass in the middle of the experimental setup (see Fig. \ref{fig:Gesamtaufbau}). In Fig.~\ref{fig:Aufstellung} the quench currents of magnets at the ends of the strings are indicated.

\begin{figure}[htp]
\centering
\includegraphics[scale=0.45]{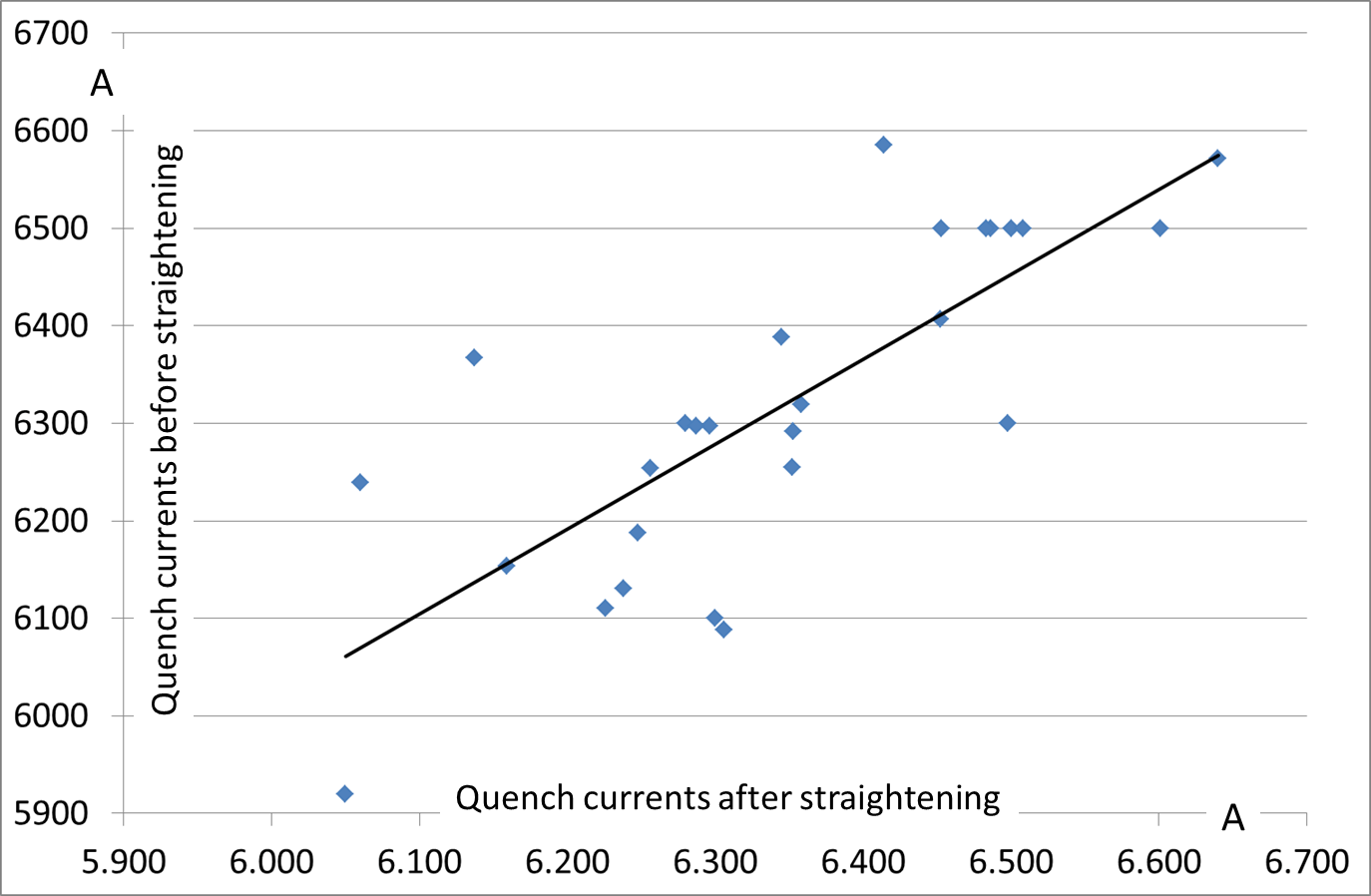}
\caption{\em \small{ Quench currents of  dipoles before straightening vs. after straightening.} } 
\label{fig:quenchvornach}
\end{figure}

 The concern, that the performance of the dipoles would be deteriorated by the straightening, did not substantiate (see Fig.~\ref{fig:quenchvornach}). The differences between the dipole quench currents, before and after straightening, are below 3\%. 

Finally, we succeeded to obtain dipoles with sufficiently large horizontal apertures and sufficiently large quench currents for two strings of 12 dipoles each, plus two spares.

The slight polygon shape of the outer vacuum vessel does not pose any problem in connecting adjacent magnets in the straight magnet strings, as verified in a test with two dipoles..


Among the selected dipoles for the experiment are 6 dipoles with a pumping connection from the outer vacuum vessel to the beam tube (see~\cite{Boehnert:1992mb}). These will be evenly distributed along the strings, allowing adequate pumping of the beam tube at room temperature.

A measurement of the additional heat loads on the 4K and 70K level, caused by the new suspensions and the pressure props, was performed. For this, the heat loads  of a dipole were measured on the test bench before and after the modifications made during the straightening procedure. However, the heat loads from the adjacent cryogenic boxes dominated the measurements and their uncertainties, preventing a determination of the additional heat load of the magnet by the modifications. The measured heat loads were the same in both cases within the uncertainties, with the heat loads of the straightened dipole being even slightly lower than that of the original dipole.  

The magnetic stray field was measured, as it may influence the detection of  regenerated photons in the ALPS II experiment. The measurement was performed on the test bench with a pickup coil at the end of a dipole about 1 m away laterally from the axis and 0.5 m below the middle of the magnet. The stray field rises with the magnet current and amounts to 0.5 Gauss at the nominal operating current of 5963 A for ALPS II.

\section{Summary}
The geometry of the HERA tunnel allows to install  2*12 HERA dipoles, at most,  as straight strings, due to the curvature of the tunnel, starting at around 100 m from the middle of hall North. Optical resonators for infrared light at a wavelength of 1064 nm,  corresponding to this length,  require a horizontal aperture at the outer ends of the experiment of 45.5 mm, for losses below $ 2*10^{-6} $ in the beam pipe.

This requirement was clearly met, as dipoles with larger apertures were selected to be positioned near the outer ends of the experiment and magnets with smaller aperture near the middle of the experiment, following the envelope of the photon beam in the optical resonators. There is about $\pm$ 2mm free space for adjustment uncertainties, without limiting the performance of the optical resonators.

By the 'brute force' straightening procedure we succeeded to obtain dipoles with sufficiently large horizontal apertures and sufficiently large quench currents for two strings of 12 dipoles each, plus two spares.

At the time of writing this report the installation of the dipole strings in the HERA tunnel is progressing.

\newpage
\section*{Acknowledgments}

Special thanks go to Gerald Meyer for most of the detailed layout of the straightening procedures and the developments of the tools.
The idea for a brute force straightening was stimulated by R\"udiger Bandelmann, when proposing the G10 transverse stabilization rods in the middle of the HERA dipole for a deformation of the cold mass.\\

\noindent The DESY survey group MEA2 with Wolf Benecke, Yvonne Imbschweiler, Nicolai Lass, and Hans Peter Lohmann made the survey of the beam pipe during the straightening of the dipoles.\\

\noindent Daniel Meissner from the DESY design office ZM1 performed the FEM calculations, necessary for the approval of the straightening procedure by the agency for pressure vessel safety (TUEV).\\

\noindent  Babette D\"obrich and Martin Berg were of great help during the assembly of the first dipoles and the functional test of the pressure props.\\

\noindent The DESY group MEA with Philipp Altmann, Clemens B\"osch, Stefan Baark, and Henrik Weitk\"amper were a great support during the assembly of the first dipoles.  Clemens B\"osch designed and fabricated the tool to transfer the position of the shield box from the original to the new suspensions. Henrik Weitk\"amper did the modifications of the shield box for the first dipoles.\\

\noindent Members of the DESY cryogenics group MKS were of great support during the straightening of the HERA dipoles: Christian Hagedorn was strongly involved in the straightening work  and in machining of parts, Olaf Fuhr also helped  in the straightening work, Oliver Paschold was involved in magnet installation and  the welding of survey marks on the vacuum vessel, Thomas T\"odten organized the magnet storage and the repainting of the dipole vacuum vessels.\\

\noindent Martin Sch\"afer and Frank Wien were very cooperative in modifying most of the shield boxes without any delay .\\

\noindent The electrical connections of the magnet to the cryogenic boxes were made by J\"urgen Eschke and Matthias Stolper in the beginning, later Olaf Sawlanski replaced Matthias Stolper.\\

\noindent The measurement of the quench current was performed by Heiner Br\"uck and Matthias Stolper for the first dipoles, later by Matthias Stolper and Olaf Sawlanski. Lothar Steffen installed a prototype quench detection system in parallel, later to be used for the ALPS II experiment.\\

\noindent Wolfgang Ratuschni from the DESY group MKK made sure that the power supply operated properly.\\

\noindent Thanks go to the operators of the central DESY cryogenics plant for the cryogenic operation of the dipoles on the test bench.\\

\noindent Bernd Petersen as group leader of the cryogenics group MKS at DESY and Detlef Sellmann as his deputy  followed the work on the test bench and supported the activities.\\

\noindent Dennis Lenz and the DESY transportation group of Uwe Eggerts managed the transportation of the dipoles and the positioning on the test bench.\\

\noindent The DESY vacuum group MVS with Thomas Kurps, Rene Ritter, Matthias Schwalger, Antonio Wagner and Sven Lederer were very supportive on the insulating vacuum system of the dipoles.\\

\noindent  Jan Kuhlman made the new CAD drawings for the HERA dipole and the modifications.\\

\noindent Bj\"orn Hager and the central DESY Workshop supplied the parts for the pressure props and the new suspensions. The first pressure props were fabricated by Uwe Packheiser.\\

\noindent We want to thank the DESY directorate for allowing, that personnel engaged in the preparation of the work and the necessary tests. \\

\noindent Finally we want to especially thank Axel Lindner-- the spokesperson of the ALPS II experiment-- for his encouragement and continuous support of this work, and careful reading of the draft.

\end{document}